\documentclass[twocolumn]{svjour3}          
\smartqed  
\usepackage{graphicx}
\usepackage{mathptmx}      

\usepackage{appendix}

\newcommand{\syst}{HybEA} 
\newcommand{\systK}{HybEA-K}
\newcommand{\systR}{HybEA-R}

\newcommand{\DOne}{D-W~(S)}
\newcommand{\DTwo}{D-W~(D)}
\newcommand{\DThree}{D-W~(SRPRS-N)}
\newcommand{\DFour}{D-W~(SRPRS-D)}
\newcommand{\DFive}{BBC-DB}

\newcommand{\DSix}{FR-EN}
\newcommand{\DSeven}{JA-EN}
\newcommand{\DEight}{ZH-EN}

\newcommand{\DNine}{ICEWS-WIKI}
\newcommand{\DTen}{ICEWS-YAGO}

\usepackage[normalem]{ulem}
\usepackage[mathscr]{eucal}
\usepackage{bm}
\usepackage{multirow}
\usepackage{placeins} 
\usepackage{soul}
\usepackage{hyperref}
\usepackage{amsmath}
\usepackage{amssymb}
\usepackage{xcolor}
\usepackage[scr=boondox,  
            cal=esstix]   
           {mathalpha}

\usepackage{subcaption} 
\usepackage[linesnumbered,ruled,vlined,algo2e]{algorithm2e}
\begin{document}

\title{\syst: Hybrid Models for Entity Alignment}


\author{Nikolaos Fanourakis         \and
        Fatia Lekbour               \and
        Guillaume Renton            \and
        Vasilis Efthymiou           \and 
        Vassilis Christophides
}


\institute{N. Fanourakis \at
              Computer Science Department, University of Crete, Greece\\
              \email{nikosfanou@csd.uoc.gr}           
           \and
           F. Lekbour \at
              ETIS, CYU University, France\\
              \email{fatia.lekbour@ensea.fr}
           \and
           G. Renton \at
              ETIS, ENSEA, France\\
              \email{guillaume.renton@ensea.fr}
            \and
            V. Efthymiou \at 
              Department of Informatics and Telematics, Harokopio University of Athens, Greece \& FORTH-ICS, Greece\\
              \email{vefthym@hua.gr}      
            \and 
            V. Christophides \at 
              ETIS, ENSEA, France\\
              \email{vassilis.Christophides@ensea.fr}
}

\date{Received: date / Accepted: date}

\maketitle

\begin{abstract}
Entity Alignment (EA) aims to detect descriptions of the same real-world entities among different Knowledge Graphs (KG). Several embedding methods have been proposed to rank potentially matching entities of two KGs according to their similarity in the embedding space. However, existing EA embedding methods are challenged by the diverse levels of structural (i.e., neighborhood entities) and semantic (e.g., entity names and literal property values) heterogeneity exhibited by real-world KGs, especially when they are spanning several domains (DBpedia, Wikidata). Existing methods either focus on one of the two heterogeneity kinds depending on the context (mono- vs multilingual). To address this limitation, we propose a flexible framework called HybEA, that is a hybrid of two models, a novel attention-based factual model,  co-trained with a state-of-the-art structural model.
Our experimental results demonstrate that HybEA outperforms 
the state-of-the-art EA systems, achieving a 16\% average relative improvement of Hits@1, ranging from 3.6\% up to 40\% in 5 monolingual datasets, with some datasets that can now be considered as \emph{solved}. We also show that HybEA outperforms state-of-the-art methods in 3 multi-lingual datasets, as well as on 2 datasets that drop the unrealistic, yet widely adopted, one-to-one assumption. Overall, HybEA outperforms all (11) baseline methods in all (3) measures and in all (10) datasets evaluated, with a statistically significant difference.

\end{abstract}

\section{Introduction}\label{sec:introduction}

Knowledge graphs (KGs) are becoming ubiquitous in a vast variety of domains, supporting applications like question answering~\cite{DBLP:conf/aaai/AhmetajEFKLO021}, entity search~\cite{DBLP:conf/kdd/0001GHHLMSSZ14}, and recommendation systems~\cite{DBLP:journals/air/TarusNM18}. Typically, KGs store both \emph{structural} and \emph{factual} information of real-world \emph{entities} (e.g., people, movies, books), under the form of subject-predicate-object triples. A crucial task when integrating knowledge from several KGs is to detect whether two KG nodes represent the same real-world entity i.e., they \emph{match}, a task known as \emph{entity alignment (EA)}~\cite{DBLP:journals/datamine/FanourakisEKC23,DBLP:journals/pvldb/SunZHWCAL20,DBLP:journals/pvldb/SuchanekAS11}, entity matching~\cite{DBLP:journals/pvldb/0001LSDT20}, or entity resolution~\cite{DBLP:series/synthesis/2015Christophides,DBLP:journals/csur/ChristophidesEP21}.

Several KG \emph{embedding} methods have been proposed for EA
~\cite{DBLP:journals/pvldb/HuoCKNHLLNLQ24,DBLP:journals/datamine/FanourakisEKC23,DBLP:journals/pvldb/SunZHWCAL20,DBLP:journals/aiopen/ZengLHLF21,DBLP:journals/corr/abs-2107-07842,DBLP:journals/tkde/WangMWG17,DBLP:conf/coling/ZhangLCCLXZ20,zhao2020experimental,DBLP:conf/dsc/JiangLG21,DBLP:conf/dsc/WangLG21,DBLP:journals/pvldb/LeoneHAGW22,DBLP:journals/vldb/ZhangTLJQ22,DBLP:conf/www/LiuHWCKD022,DBLP:journals/corr/abs-2205-08777}, as they can potentially mitigate the symbolic, linguistic and schematic heterogeneity of independently created KGs. The idea is to represent the nodes (entities) and edges (relations or properties) of KGs in a low-dimensional space, in a way that similar entities in the original KG are placed close to each other in the embedding space. 
In this respect, they leverage the \emph{structural} (i.e., entity neighborhood) and the \emph{factual} (i.e., entity names/identifiers, literal values) information  of entity descriptions. Given that different KGs emphasize on different aspects of entities, one of the major EA challenges is to cope with heterogeneous entity descriptions~\cite{DBLP:series/synthesis/2015Christophides}.

\begin{figure} [t]
    \centering
    \includegraphics[width=0.50\textwidth]{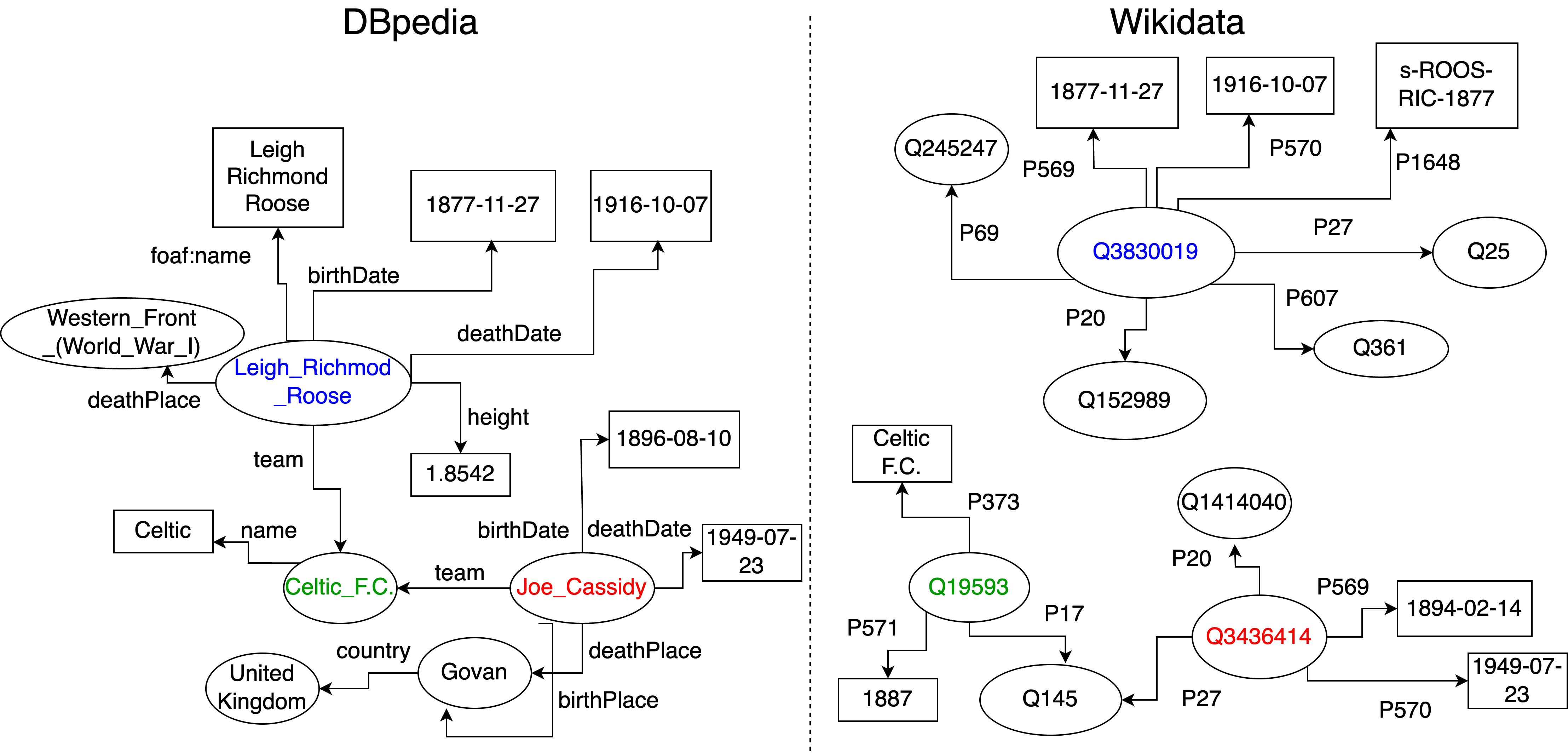}
    \caption{Motivating example from a real dataset (\DOne). Entities (nodes) with the same color are matching entities.}
    \label{fig:motivating_example}

\end{figure}

\begin{example}
Figure~\ref{fig:motivating_example} depicts a small subset of aligned entities in two real-world KGs, DBpedia (left) and Wikidata (right) that exhibit both high semantic (e.g., entity and property names) and structural (e.g., entity neighborhood) heterogeneity and which are challenging for most state-of-the-art EA methods for various reasons. Let's consider \textcolor{blue}{\textit{Leigh\_\\Richmod\_Roose}} in DBpedia and \textcolor{blue}{\textit{Q3830019}} in Wikidata, which should be aligned. Apart from the names, the factual information related to these entities is relatively similar: \textit{birthDate} and \textit{deathDate} in DBpedia (resp. \textit{P569} and \textit{P570} in Wikidata) have the same information, crucial to find the correct alignment. However, these entities are structurally very different. They don't have any common neighbor nor the same number of neighbors. We call this structural heterogeneity, and methods that are based on structural information, such as Knowformer~\cite{10092525} or RREA~\cite{DBLP:conf/cikm/MaoWXWL20}, may struggle with this example.  

If we consider now \textcolor{red}{Joe\_Cassidy} from DBpedia, which should be aligned with \textcolor{red}{Q3436414} from Wikidata, we observe semantic heterogeneity on the attributes: the values of \textit{birthDate} and \textit{P569} are different, which challenges methods that are based on such information, like COTSAE~\cite{DBLP:conf/aaai/YangLZWX20}, ZeroEA~\cite{DBLP:journals/pvldb/HuoCKNHLLNLQ24} or BERT-INT~\cite{DBLP:conf/ijcai/Tang0C00L20}. However, when the factual information is not consistent, the structural information may help: still with \textcolor{red}{Joe\_Cassidy}, we can see see that it is a direct neighbor of \textit{Govan} (\textit{Q1414040} in Wikidata), and close to \textit{Celtic\_F.C}. Structural methods which use such information can thus manage to align these entities.

\end{example}


As real KGs exhibit different levels of structural and semantic heterogeneity we need to consider both entity relations and facts during alignment in an adaptive way, without prioritizing one or the other. Nevertheless, existing embedding models for EA focus mainly on either structural (e.g., Knowformer~\cite{10092525}, RREA~\cite{DBLP:conf/cikm/MaoWXWL20}, SelfKG~\cite{DBLP:conf/www/LiuHWCKD022}) or semantic (e.g., COTSAE~\cite{DBLP:conf/aaai/YangLZWX20}, ZeroEA~\cite{DBLP:journals/pvldb/HuoCKNHLLNLQ24}, BERT-INT~\cite{DBLP:conf/ijcai/Tang0C00L20}) heterogeneity and they are usually evaluated with datasets that do not highly exhibit both heterogeneity forms. In this paper, we introduce a novel EA framework that can dynamically adapt to the different levels of semantic and structural heterogeneity exhibited by real KGs in a semi-supervised way. More precisely, \syst\ enables to form a \emph{hybrid} model that exploits both the entity relations and facts.  The \syst\ factual component is capable of weighting the contribution of each property in the prediction of matching entity pairs using a \emph{new attention-based model}. Then, several hybrids can be formed by plugging different structural models that exploit the relations connecting neighbor entities in KGs at the level of triples (e.g., Knowformer~\cite{10092525}) or graphs (RREA~\cite{DBLP:conf/cikm/MaoWXWL20}).
As each component model predicts its own entity alignments, we propose \emph{a new semi-supervised framework}, where at every iteration, some high-confidence matching pairs detected by each model are added to the training set used for the subsequent co-training of both models.
In summary, the contributions of this work are the following: 
\begin{itemize}
    \item We introduce a novel semi-supervised, co-training EA framework, \syst, allowing to create hybrids of a novel factual model with state-of-the-art structural models. High-confidence predictions are then suggested for retraining both models in subsequent iterations by adapting to the heterogeneity characteristics of different datasets.
    \item We provide an in-depth analysis of ten monolingual and multilingual datasets to reveal the different levels of structural and semantic heterogeneity exhibited by real KGs. Using a wide range of metrics, we group existing datasets into 6 clusters that have only been partially considered in previous experimental evaluations~\cite{DBLP:journals/pvldb/SunZHWCAL20,DBLP:journals/datamine/FanourakisEKC23,DBLP:conf/www/JiangXSWSSSLGS24}. 
    \item We experimentally demonstrate that \syst\ \emph{outperforms} eight state-of-the-art EA systems (e.g., BERT-INT, RREA, COTSAE, PARIS+), achieving a 16\% average relative improvement of Hits@1, ranging from 3.6\% up to 40\% in all 5 monolingual datasets, with some datasets that can now be considered as \emph{solved}. Moreover, \syst\ outperforms all other baselines in 3 widely used multilingual datasets. Finally, \syst\ also presents impressive results on 2 mono-lingual datasets that drop the one-to-one assumption. \emph{Overall, \syst\ outperforms all (11) baseline methods in all (3) measures and in all (10) datasets evaluated, with a statistically significant difference}.
    \item The source code and the datasets used are publicly available: \url{https://github.com/fanourakis/HybEA.git}
    \end{itemize}

The remainder of the paper is organized as follows. In Section~\ref{sec:related_works}, we overview related works and compare them to our approach. In Section~\ref{sec:methodology}, we present our methodology. Section ~\ref{sec:dataset_analysis} we analyse the levels of structural and semantic heterogeneity exhibited by popular datasets used in experimental evaluations of EA systems. The performance of \syst\  against 11 baseline methods in these datasets is detailed in Section~\ref{sec:experiments}.  We conclude and present directions for future work in Section~\ref{sec:conclusion}.

\section{Related Work}\label{sec:related_works}


Despite the many names and settings of EA problem, most related works can be broadly classified with respect to their focus on the  structural information (e.g., entity relationships through foreign keys, dependency graphs, or relations/edges in KGs) or on the factual information of entities (e.g., their literal values). In the sequel, we will position \syst\ in the context of these two general families of EA systems. 

\textbf{Structure-based approaches. }
Knowing that two candidate matches have many matching neighbors, even better, through semantically similar edges, provides strong evidence that those candidates are also matches; especially when the involved edges represent functional relations~\cite{DBLP:journals/pvldb/SuchanekAS11}). 

In this respect, there are several EA methods that learn vector representations (embeddings) of nodes (entities), encoding structurally similar nodes close in the vector space, while dissimilar far. Earlier structural methods usually rely on TransE~\cite{DBLP:conf/nips/BordesUGWY13} interpreting entity relations as translations operating on the node embeddings. MTransE~\cite{DBLP:conf/ijcai/ChenTYZ17} encodes the entities of the two KGs 
in different vector spaces using TransE, and then exploits their local structure before transitioning between them. IPTransE~\cite{DBLP:conf/ijcai/ZhuXLS17} and BootEA~\cite{DBLP:conf/ijcai/SunHZQ18} enrich the training set with new discovered matching entities, based on their structural similarity.

Moreover, several Graph Neural Networks (GNNs) have been proposed to account a wider range of each node neighborhood, than translation-based methods. However, this comes with the risk of incorporating noisy or even irrelevant to the task information to the node embeddings, from distant neighbors. In this respect, MuGNN~\cite{DBLP:journals/corr/abs-1908-09898} and AliNet~\cite{DBLP:journals/corr/abs-1911-08936} adopt an attention mechanism to weigh more important neighbors. PipEA~\cite{DBLP:journals/corr/abs-2402-03025} modifies GNNs propagation strategy in order to allow the information to also pass from one graph to the other and thus to compute a similarity measure between subgraphs of both KGs. To tackle structural heterogeneity, RePS \cite{surisetty2022reps} considers a positional encoding according to the distance of each node to a set of anchors, with the idea that rarest relations are the most informative. RREA \cite{DBLP:conf/cikm/MaoWXWL20} constraints the GNNs transformation matrix to be orthogonal and hence preserve shape similarity. It also proposes a semi-supervised setting that adds to the training set mutually nearest neighbors, while it also adopts an attention mechanism for filtering the information of noisy distant neighbors. RDGCN \cite{DBLP:conf/ijcai/WuLF0Y019} employs a variant of graph convolution, Dual-Primal GCN, to capture entity neighborhood structure using features of relationship types as well as entity names.

Finally, there also exist a few structural methods that rely neither on translational models, nor on GNNs. Knowformer~\cite{10092525} exploits Transformers to create entity embeddings using individual relation triplets of KGs. To integrate structural information such as direction and local information, Transformers are extended with a position-aware structural component, called relational composition. PARIS+ \cite{DBLP:journals/pvldb/LeoneHAGW22,DBLP:journals/pvldb/SuchanekAS11} relies on a probabilistic method to align entities, by statistically estimating matching probabilities.

Structured-based methods rely on the rather unlealistic assumption that equivalent entities from different KGs possess similar neighboring KG structures (and in turn similar embeddings). Hence, they leverage the seed entity pairs as anchors and progressively project individual KG embeddings into a unified space through training, resulting in the unified entity representations~\cite{Zeng23}. Compared to this family of EA methods, \syst\ is able to adapt to different degrees of structural and semantic heterogeneity exhibited by mono- and multilingual KGs (see Table~\ref{tab:reciprocity_rrea}). 



\textbf{Joint approaches.}
As the number of attributes significantly exceeds that of relations in real KGs (see Table~\ref{tab:datasets}), some methods jointly consider the factual and structural information of entities. Although they have been empirically proven to be highly effective~\cite{DBLP:journals/datamine/FanourakisEKC23}, their performance significantly drops in KGs exhibiting high heterogeneity at both structural and factual levels. 

JAPE~\cite{DBLP:journals/corr/abs-1708-05045} combines structure embedding and attribute embedding to match entities in different KGs, using TransE and Skip-Gram, respectively. MultiKE~\cite{DBLP:conf/ijcai/ZhangSHCGQ19} considers three views of KGs, distinguishing the name, relations and attributes of entities. These views are jointly learned using a loss function that weighs all three equally. This method assume high homogeneity in both attribute names and values, thus we expect MultiKE to perform better in datasets originating from the same or similar data sources (e.g., Wikipedia-based, or synthetic KGs). On the other hand, KDCoE~\cite{DBLP:conf/ijcai/ChenTCSZ18} is a semi-supervised method that co-trains two embedding models, one on the structural part of entities (MTransE~\cite{DBLP:conf/ijcai/ChenTYZ17}) and the other on the textual descriptions of the entities. Similarly, COTSAE~\cite{DBLP:conf/aaai/YangLZWX20} co-trains a structural model based on TransE~\cite{DBLP:conf/nips/BordesUGWY13} along with a general attention-based factual model and employs a bipartite graph matching algorithm to find the global optimum result. COTSAE's attention mechanism can capture only short-term dependencies, and thus not all attributes of KGs are exploited for learning the attribute attentions of an entity. In addition, the new matches of the semi-supervised COTSAE rely solely on the refined structured entity embeddings, which are usually false positives (especially when structural heterogeneity is high), introducing noise to the model, resulting to low adaptability of the method in terms of structural and semantic heterogeneity.

\textbf{Language Model-based approaches.} Pre-trained language Models (PLM) have been employed to cope with the semantic heterogeneity of KGs. AttrGNN~\cite{DBLP:journals/corr/abs-2010-03249} leverages masked language models like BERT~\cite{DBLP:conf/naacl/DevlinCLT19} to learn attribute value and name embeddings (literal, digital and name channel) and multi-layer GCN for learning the structure embeddings (structure channel). It also uses either average pooling or LS-SVM~\cite{DBLP:journals/npl/SuykensV99} for ensembling the information provided by the aforementioned channels, as well as an attribute attention mechanism based on GAT. On the other hand, BERT-INT~\cite{DBLP:conf/ijcai/Tang0C00L20} leverages masked language models to contextualize literal descriptions of entities in embeddings. More precisely, BERT-INT fine-tunes BERT to encode entity names, textual descriptions, and attribute values of entities, performing pairwise neighbor-view and attribute-view interactions to obtain entity matching scores. To reduce the number of unnamed nodes for which the method considers only random embeddings, BERT-INT relies on additional textual information not necessarily included in the original multilingual datasets~\cite{DBLP:conf/acl/TangZL23}. ZeroEA~\cite{DBLP:journals/pvldb/HuoCKNHLLNLQ24} also leverages BERT over entire entity neighborhoods along with a natural language translator to obtain an embedding that is directly used to compute the similarity scores. 
Simple-HHEA rely on random walks~\cite{DBLP:journals/inffus/WangHWWZW23} to encode the structure of entities and on BERT to initialize the entity embeddings using their name. It also uses an entity time encoder for exploiting the temporal information of KGs (if it is available). Finally, to overcome the scarcity of manually labeled entity pairs, SelfKG \cite{DBLP:conf/www/LiuHWCKD022} adapts the NCE loss for self-supervision with self-negative sampling along a GAT and LaBSE for initializing the entity embeddings with their names. As it is demonstrated later in the experimental results (see Table~\ref{tab:effectiveness}), despite the power of the PLMs, PLM-based methods require low semantic heterogeneity to provide satisfactory effectiveness.

Generative large language models (LLM) like GPT-4\footnote{\url{openai.com/index/gpt-4}} or Llama\footnote{\url{llama.meta.com}} have been also used in EA tasks due to their improved text generation and fine-grain instrumentation capabilities~\cite{chen2024entity,yang2024two,DBLP:journals/pvldb/NarayanCOR22,jiang2024unlocking}. For example, LLM4EA~\cite{chen2024entity} aims at reducing the quantity of labeled data required for training, by generating pseudo-labels of entity pairs, along with a label refiner that removes incompatible labels. Then, an embedding-based EA model is trained over the refined labels, to learn structure-aware embeddings and compute a matching score for each entity pair. On the other hand, ChatEA~\cite{DBLP:conf/acl/JiangSSXLLGSW24} translates the KG structure to a code format that is more convenient to take benefit of LLMs reasoning capabilities, while Simple-HHEA (in the entity feature pre-processing step) exploits LLMs in candidate matches selection. Similarly, LLMEA uses relation-aware GAT for candidate selection and LLMs for the matching prediction, while LLM-Align~\cite{DBLP:journals/corr/abs-2412-04690} uses heuristic methods to select important attributes and relations of entities, in order the corresponding attribute and relation triples to be fed in the LLM. The last two preliminary works do not provided code, while they have not been published, yet. Clearly, Language Model-based EA methods incur multiple LLM queries whose cost remains relatively high\footnote{\url{docs.swarms.world/en/latest/guides/pricing}}. 





\section{Methodology}\label{sec:methodology}
In this section, we present our methodology, starting with a brief description of the problem setting and an overview of our approach. Then, we provide details about every component of our approach. 

\subsection{Problem Setting}\label{ssec:problem}
Following the typical notation~\cite{DBLP:conf/ijcai/ZhangSHCGQ19,DBLP:conf/emnlp/WangYY20,DBLP:journals/datamine/FanourakisEKC23}, we assume that entities are described in KGs by a collection of triples $\left<h,r,t\right>$, whose head $h$ is always an entity, and tail $t$ may be either an entity, in which case we call $r$ a \emph{relation} (and $\left<h,r,t\right>$ a \emph{relation triple}), or a literal (e.g., number, date, string), in which case we call $r$ an \emph{attribute} (and $\left<h,r,t\right>$ an \emph{attribute triple}). 

We represent a knowledge graph as $KG =$ ($E$, $R$, $A$, $L$, $X$, $Y)$, where $E$ is a set of entities (e.g., Joe\_Cassidy, Celtic\_F.C.), $R$ is a set of relations (e.g., team), $A$ is a set of attributes (e.g., birthDate), $L$ is a set of literals (e.g., ``1896-08-10''), and $X \subseteq(E \times R \times E)$ and $Y \subseteq(E \times A \times L)$ are the sets of relation (e.g., $\left<Joe\_Cassidy, team, Celtic\_F.C.\right>$) and attribute triples (e.g., $\left<Joe\_Cassidy, birthDate, ``1896-08-10''\right>$) of the KG, respectively. Given a source $KG_1 = (E_1, R_1, A_1, L_1, X_1, Y_1)$ and a target $KG_2$ = $(E_2, R_2, A_2$, $L_2, X_2, Y_2)$, the task of entity alignment is to find pairs of matching entities $M = \{(e_i,e_j)\in E_1 \times E_2 \mid e_i \equiv e_j \}$, where ``$\equiv$'' denotes the equivalence relationship \cite{DBLP:conf/ijcai/ZhangSHCGQ19,DBLP:conf/emnlp/WangYY20}. A subset $S \subseteq M$ of the matching pairs may be used as a seed alignment for training, along with a set $N \subseteq  (E_1 \times E_2) \setminus M$ of non-matching pairs, called negative samples. By convention, the one-to-one assumption constraint is generally considered~\cite{DBLP:conf/ijcai/SunHZQ18,DBLP:conf/aaai/YangLZWX20,DBLP:journals/corr/ZhangRG14}, where every entity in $E_1$ should be matched to exactly one entity in $E_2$: $\forall e_i \in E_1\;  \left(\exists e_j \in E_2: (e_i, e_j) \in M\right) \wedge \left(\nexists e_j' \in E_2: (e_i, e_j') \in M\right)$. \syst\ is not constrained by the one-to-one assumption; we show experimental results (Section \ref{sec:experiments}) on both datasets that follow and datasets that do not follow this constraint.



\vspace{-.5cm}
\subsection{\syst ~Overview}\label{ssec:overview}


The architecture of \syst\ involves a \emph{factual component} and a \emph{structural component}. The factual component, built on an attention mechanism, is responsible for gathering matching evidence from attribute triples. The structural component, on the other hand, is responsible for gathering matching evidence originating from structural information of the graph. Our model is constructed around our factual component, in which we can plug different structural models, in our case Knowformer and RREA, which offers a flexibility in choosing the desired structural component according to the dataset specificity.

Our framework is semi-supervised; it runs in cycles, with each cycle involving the co-training of the factual component and the structural component. Each component identifies a set of \emph{new matching pairs} which are added into the training set and used for training the next component. Other than exchanging pairs that each model considers as matches, used for co-training both models, the two models are otherwise independent from each other. The execution stops when no new matching pairs are identified by at least one component at the end of a cycle, or a maximum number of cycles has been reached.

The new matching pairs, regardless of the component from which they originate, are considered as high-confidence pairs, so they are also part of \syst's returned matches. For the remaining entities, not belonging to high-confidence matching pairs, we use the suggestions provided by a bipartite graph matching algorithm ran on the similarity matrix of the component that was executed last. Typically, the Best Match algorithm is utilized in EA methods, i.e., a ranked list of $KG_2$ entities sorted in decreasing match likelihood for every $KG_1$ entity is returned, so we also follow this approach. More options are discussed in~\cite{DBLP:journals/vldb/PapadakisETHC23,DBLP:journals/csur/ChristophidesEP21}. 


The high-level algorithm of HybEA, without the specific algorithmic decisions on the embedding initialization, the factual and the structural model, as well as the bi-partitie graph matching algorithm, which are all described next in detail, is presented in Algorithm~\ref{algo:HybEA}, while the detection of high-confidence pairs is presented in Algorithm~\ref{algo:reciprocity}. 

\begin{algorithm2e}[t]
\KwIn{ $KG_1 = (E_1, R_1, A_1, L_1, X_1, Y_1)$, $KG_2$ = $(E_2, R_2, A_2$, $L_2, X_2, Y_2)$, disjoint subsets of matching pairs $S_{train}$, $S_{val}$, $S_{test} \subseteq M$} 
\KwOut{Matching pairs $M' \subseteq E_1 \times E_2$}

$M' \gets \emptyset$ \tcp*{set of matching pairs}
$SMat[E_1][E_2]$ \tcp*{similarity matrix $E_1 \times E_2$}

$\vec{E_1} \gets embed(KG_1)$ \tcp*{embedding initialization}
$\vec{E_2} \gets embed(KG_2)$ \tcp*{embedding initialization}

\For{$i$ in [1..maxCycles]}{
    \tcc{train the factual model on  the (enriched) training set, update $SMat$ and $M'$}
    $fm \gets$ train\&validate\_fact\_model($S_{train} \cup M'$, $S_{val}, \vec{E_1}, \vec{E_2})$\\
    $SMat \gets fm.run(S_{test})$ \tcp*{sim. matrix from fm}
    $M_{fm} \gets reciprocity(SMat)$  \tcp*{new pairs from fm}
    $M' \gets M' \cup M_{fm}$ \\
    
    \BlankLine
    \tcc{do the same for the structural model}
    $sm \gets$ train\&validate\_str\_model($S_{train} \cup M'$, $S_{val},  \vec{E_1}, \vec{E_2})$\\
    $SMat \gets sm.run(S_{test})$ \tcp*{sim. matrix from sm}
    $M_{sm} \gets reciprocity(SMat)$ \tcp*{new pairs from sm}
    $M' \gets M' \cup M_{sm}$ \\
    

}

$M'_{rest} \gets BipartiteGraphMatching(SMat)$\\

$M'' \gets \{(e_i, e_j) \in M'_{rest} | \nexists e_k, (e_i, e_k) \in M'\}$ \\

$M' \gets M' \cup M''$ \tcp*{reciprocal matches $M'$ and non-reciprocal matches $M''$ from the last model}

\textbf{return} $M'$

\caption{HybEA high-level algorithm.}
\label{algo:HybEA}
\end{algorithm2e}

\begin{algorithm2e}[!htb]
\KwIn{Similarity Matrix $SM_{|E_1| \times |E_2|}$}
\KwOut{(Reciprocally) Matching pairs $RM \subseteq E_1 \times E_2$}

$RM \gets \emptyset$\\
\ForEach{$j \in 1..|E_2|$}{%
    $TopCandE_2[j] \gets SM^T[j].indexOf(SM^T[j].max())$
}

\ForEach{$i \in 1..|E_1|$}{%
    $j \gets SM[i].indexOf(SM[i].max())$ \\
    \If{$TopCandE_2[j] = i$}{
        $RM \gets RM \cup \{(e_i, e_j)\}$ \tcp*{$e_i \in E_1$ and $e_j \in E_2$}
    }
}

\textbf{return} $RM$

\caption{Reciprocity filter.}
\label{algo:reciprocity}
\end{algorithm2e}


\subsection{Exploiting Factual Information}\label{ssec:factual}

Let $A_j = (a^j_1, a^j_2, ..., a^j_n)$ be the totally ordered set of attributes
in $KG_j$.
We want to identify which attributes from $A_1$ and $A_2$ 
are more important for aligning $KG_1$ and $KG_2$. Thus, we want to learn the attention weights $\mathcal{A}_j = (\mathscr{a}^j_1, \mathscr{a}^j_2, \ldots, \mathscr{a}^j_n)$ corresponding to the attributes $A_j = (a^j_1, a^j_2, ..., a^j_n)$, using the attributes of the entire $KG_j$ along with their literals (values).

Let $\boldsymbol{A} = (\vec{a_1}, \vec{a_2}, ..., \vec{a_n})$ be the attribute embeddings of the attributes $A = (a_1, a_2, \ldots, a_n)$, and let $\boldsymbol{V} = (\vec{u_1}, \vec{u_2}, ..., \vec{u_n})$ be the literal (value) embeddings corresponding to the attributes in $A$, where $v_i \subseteq L$ are literal values. Attention weights are computed by our factual model using 
the scaled dot-product attention~\cite{DBLP:conf/nips/VaswaniSPUJGKP17}, as follows:
\vspace{-0.3cm}
\begin{equation}\label{eq:attention}
    Attention(Q,K,V) = \text{softmax}(\frac{QK^T}{\sqrt{d_k}})V,
\end{equation}
where $Q$ (query) are the attribute type embeddings of an entity $e$ with $m$ attributes, $K$ (key) corresponds to the attribute type embeddings~$\boldsymbol{A}$ of the entire $KG_j$ with $A_j$ attributes, $V$ (value) corresponds to the literal (values) embeddings of~$\boldsymbol{A}$, and $d_k$ is the dimension of the embeddings, that are used as scaled factor to avoid pushing softmax function to regions where it has extremely small gradients~\cite{DBLP:conf/nips/VaswaniSPUJGKP17}.

Having computed the attribute attention weights $\mathcal{A}$, 
we now compute the embedding of an entity $e$, as 
\vspace{-0.3cm}
\begin{equation}
    \vec{e} =  \sum_{(e,a_i,u_i)\in Y} \mathscr{a}_i\vec{u_i}.
\end{equation}


We employ a pseudo-Siamese network architecture, which has shown good results in~\cite{DBLP:conf/aaai/YangLZWX20}, to compute different attention weights $\mathcal{A}_1$ and $\mathcal{A}_2$ for $KG_1$ and $KG_2$, respectively, while we compute the literal value embeddings $\boldsymbol{V}$ of both networks using the same pre-trained sentence embeddings model (i.e., Sentence-BERT~\cite{DBLP:conf/emnlp/ReimersG19}).

Finally, to account factual information, we compute the distance of two entities $e_1 \in E_1$ and $e_2 \in E_2$, as the Euclidean distance of their embeddings 
$d(e_1, e_2) = ||\vec{e}_1 - \vec{e}_2||_2$, and
we use the contrastive loss function
\vspace{-0.3cm}
\begin{equation}
    \mathcal{L}_a = (1-\alpha) \sum_{(e_1,e_2) \in M} d(e_1, e_2) +  \alpha \sum_{(e_1',e_2') \in N}   [\lambda - d(e_1, e_2)]_+,
\end{equation}
where $0 \leq \alpha \leq 1$ is a hyper-parameter for weighting the negative samples over the positive samples, $\lambda$ is a margin used for the distance of negative samples (the distance of the negative samples is expected to be larger than $\lambda$), and $[·]_+ = max(0, ·)$.

The training and validation of the factual model is abstractly defined in Line 6 of Algorithm~\ref{algo:HybEA}, using the initial training set $S_{train}$ enriched with any newly discovered pairs $M'$ as training and the initial validation set $S_{val}$ for validation.
The newly trained factual model $fm$ returns a similarity matrix $SMat$.

\begin{figure} [t]
    \centering
    \includegraphics[width=0.50\textwidth]{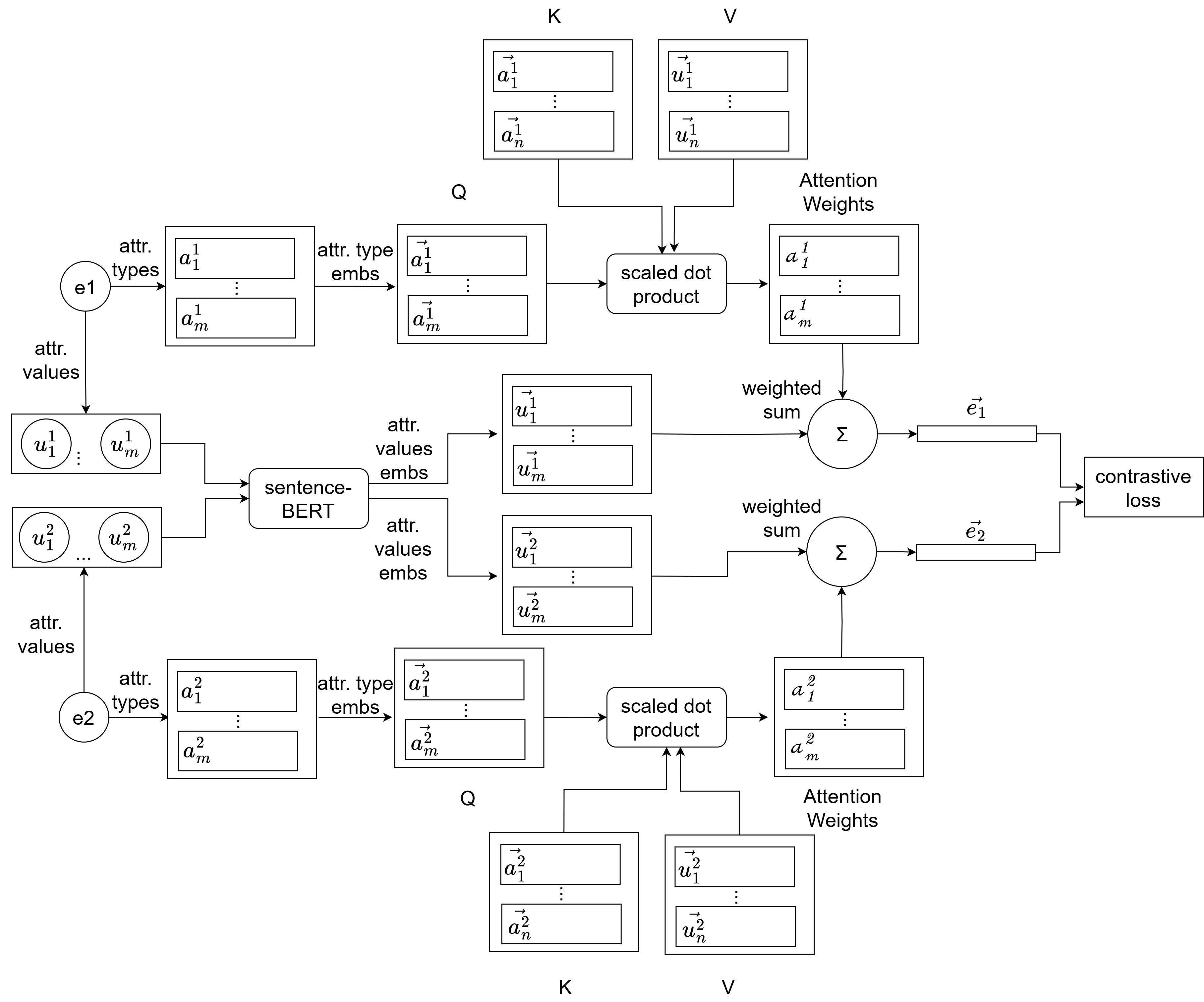}
    \caption{Factual model architecture.}
    \label{fig:fact_architecture}
\end{figure}

\subsection{Exploiting Structural Information}\label{ssec:structural}



Our method is built around the factual component and our structural component works as a plug-and-play component, for which we can use different structure-based models.
For the choice of the structural component, two methods were selected: Knowformer and RREA.


Those two approaches are based on different principles: Knowformer focuses on local relation, while RREA exploits the global structure of the graph. First, Knowformer was chosen because it allows good management of the expressiveness of relationships. This method focuses directly on triplet-based structure. In fact, it introduces relational composition using translation methods, which allow encoding position in triplets and distinguishing every entity, whether it is a subject or an object. Its use of attention mechanisms strengthens its ability to capture different local information. 

The second method that was chosen is RREA, an attention GNN-based method. GNN propagates information through the graph and allows an explicit knowledge of the graph that is not only locally based, contrary to Knowformer. Through relational reflection transformations, an orthogonal constraint, RREA effectively propagates information across the graph, while preserving spatial integrity.  
It does not differentiate specific relation as precisely as Knowformer does, but instead provides a more global context of the graph.

Furthermore, we modified Knowformer by integrating Sentence-BERT embeddings, leading to a variant that we refer to as \systK\ (w/o struct.) in \systK. However, we did not apply the same modification to RREA, as shown in Table~\ref{tab:performance_hybEAR} of Appendix \ref{Appendix_reciprocity}, since initializing the model with either Sentence-BERT or RREA embeddings does not lead to significant performance differences in our model.

Finally, by testing with these two methods, we demonstrate the flexibility of our approach. RREA is implemented in TensorFlow, while Knowformer is built in PyTorch, showing that our framework is adaptable to various models and allows easy interchangeability of the structural component.

\subsection{Semi-supervision}


In order to identify high-confidence matching pairs, we employ a reciprocity filter~\cite{DBLP:conf/edbt/Efthymiou0SC19,DBLP:conf/cikm/MaoWXWL20}, briefly described in Algorithm~\ref{algo:reciprocity}. This reciprocity filter essentially returns only those entity pairs for which both entities have the other entity as their most probable match. In other words, it keeps only those pairs $(e_i,e_j)$, with $e_i \in E_1$, $e_j \in E_2$,  for which $e_j$ is the top-1 candidate for $e_i$ (Lines 2-3), and $e_i$ is the top-1 candidate for $e_j$ (Lines 4-7).

Our novel inductive semi-supervised architecture operates as follows. We incrementally enrich
the training set with the entity pairs identified by the reciprocity filter for co-training both models (Lines 9 and 13 in Algorithm~\ref{algo:HybEA}). 
We also consider those pairs as part of the final matching pairs on which we evaluate the effectiveness of our system (Line 16 in Algorithm~\ref{algo:HybEA}). For the entities that do not fall under this reciprocity filter, we return the ranking of candidates produced by the last model that we trained (Lines 14-15 in Algorithm~\ref{algo:HybEA}). The order of executing the components, i.e., structural model first vs factual model first, is important. We experiment with both pipelines and analyze their results in the next section. Finally, we employ an early stopping mechanism, as well as a maximum number of semi-supervision cycles ($maxCycles$ in Algorithm~\ref{algo:HybEA}).

\subsection{Bipartite Graph Matching}

In the final step, for entities that have not yet been matched by the reciprocity filter, we employ a typical bipartite matching process. The standard algorithm followed by the majority, if not all, of existing works is the Best Match algorithm~\cite{DBLP:journals/vldb/PapadakisETHC23}, i.e., for every node in $KG_1$, suggest as its match the most similar entity from $KG_2$, according to the Cosine similarity of their embeddings. In \syst, we employ the embeddings that were generated by the model that was trained last. This typical approach works best under the one-to-one matching assumption, which is rather unrealistic. However, we keep this approach as followed by the literature, in order to make a direct and fair comparison to existing works. In practice, our semi-supervised algorithm can be considered as an inductive, co-training algorithm.


\begin{table*}
\centering
\caption{Statistics per dataset. As reported in Section~\ref{ssec:problem}, 
$|E|$: number of entities, 
$|X|$: relation triples, 
$|Y|$: attribute triples, 
$|R|$: relations, 
$|A|$: attributes. $wccR$ and $maxCS$ are defined in~\cite{DBLP:conf/esws/FanourakisECKPS23} as the ratio of number of weakly connected components over $|E|$, and the size of the biggest weakly connected component over $|E|$, respectively. See Section~\ref{ssec:problem} for more details on the notation.}
\label{tab:datasets}
\begin{tabular}{lrrrrrrrrr} 
\textbf{Dataset}            
& $|E|$
& $|X|$
& $|Y|$ 
& $|R|$ 
& $|A|$
& $maxCS$
& $wccR$
& nodes w/o name
& train/val/test \\ 
\hline
\textbf{\DOne} & & & & & & & & &  3,000/1,500/10,500 \\
$KG_1$ & 15,000 & 38,265 & 68,258  & 248 & 342 & 0.95 & 0.01 & 0 & \\ 
$KG_2$ & 15,000 & 42,746 & 138,246 & 169 & 649 & 0.93 & 0.02 & 3,707 & \\ 
\hline
\textbf{\DTwo} & & & & & & & & & 3,000/1,500/10,500 \\ 
$KG_1$ & 15,000 & 73,983 & 66,813  & 167 & 175 & 0.99 & 0.0006 & 0 & \\ 
$KG_2$ & 15,000 & 83,365 & 175,686 & 121 & 457 & 0.99 & 0.001 & 2,955 & \\ 
\hline
\textbf{\DThree} & & & & & & & & & 3,000/1,500/10,500 \\
$KG_1$ & 15,000 & 38,421 & 71,957  & 253 & 363 & 0.98 & 0.005 & 0 & \\ 
$KG_2$ & 15,000 & 40,159 & 136,315 & 144 & 652 & 0.99 & 0.003 & 6,581 & \\ 
\hline
\textbf{\DFour} & & & & & & & & &  3,000/1,500/10,500 \\
$KG_1$ & 15,000 & 68,598 & 62,636  & 220 & 256 & 0.99 & 0.002 & 0 & \\ 
$KG_2$ & 15,000 & 75,465 & 184,332 & 135 & 531 & 0.99 & 0.002 & 5,096 & \\ 
\hline
\textbf{\DFive} & & & & & & & & & 1,879/939/6,578 \\
$KG_1$ & 9,395 & 15,478 & 18,165  &  9 &   4 & 0.31 & 0.18 & 46 & \\ 
$KG_2$ & 9,395 & 45,561 & 149,720 & 98 & 723 & 0.78 & 0.07 &  0 & \\ 
\hline

\textbf{\DSix} & & & & & & & & & 3,000/1,500/10,500  \\
$KG_1$ & 19,661 & 105,998 & 528,665 & 903 & 4,547 & 0.99 & 0.0007 & 0 & \\ 
$KG_2$ & 19,993 & 115,722 & 576,543 & 1,208 & 6,422 & 0.99 & 0.0008 & 0 & \\ 
\hline

\textbf{\DSeven} & & & & & & & & & 3,000/1,500/10,500  \\
$KG_1$ & 19,814 & 77,214 & 354,619 & 1,299 & 5,882 & 0.98 & 0.001 & 9,429 & \\ 
$KG_2$ & 19,780 & 93,484 & 497,230 & 1,153 & 6,066 & 0.99 & 0.0008 & 0 & \\ 
\hline

\textbf{\DEight} & & & & & & & & & 3,000/1,500/10,500  \\
$KG_1$ & 19,388 & 70,414 & 379,684 & 1,701 & 8,113 & 0.98 & 0.002 & 17,280 & \\ 
$KG_2$ & 19,572 & 95,142 & 567,755 & 1,323 & 7,173 & 0.99 & 0.0009 & 0 & \\ 
\hline

\textbf{ICEWS-WIKI} & & & & & & &  & & 1,012/506/3,540  \\
$KG_1$ & 11,047 & 3,527,881 & 11,047 & 272 & 1 & 0.99 & 0.0002 & 0 & \\ 
$KG_2$ & 15,831 & 198,257 & 15,831 & 226 & 1 & 1.0 & 0.00006 & 0 & \\ 
\hline

\textbf{ICEWS-YAGO} &  &  & & & & & & & 5,142/506/13,176  \\
$KG_1$ & 26,863 & 4,192,555 & 26,863 & 272 & 1 & 0.99 & 0.0002 & 0 & \\ 
$KG_2$ & 22,555 & 107,118 & 22,555 & 41 & 1 & 0.99 & 0.001 & 0 &\\ 
\hline

\end{tabular}
\end{table*}


\section{Heterogeneity Analysis of the Datasets}
\label{sec:dataset_analysis}
In this section, we provide a detailed analysis of the different datasets we used, under the heterogeneity scope. For this purpose, after a short presentation of the different datasets used, we propose different metrics that exhibit the different levels of heterogeneity, both structurally and semantically. 

\subsection{Datasets}\label{ssec:datasets}
In this section, we present the datasets used in our experiments, and describe their main features,
as shown in Table~\ref{tab:datasets}.

\begin{itemize}
    \item \textbf{\DOne} and \textbf{\DTwo}~\cite{DBLP:journals/pvldb/SunZHWCAL20} are the sparse and dense versions, respectively, of a dataset that was constructed from DBpedia and Wikidata KGs, describing actors, musicians, writers, films, songs, cities, football players and football teams. The two variations help in detecting the influence of KG density in EA methods. These datasets also have both low entity name similarity and a low number of entities with long textual descriptions~\cite{DBLP:journals/datamine/FanourakisEKC23}, undermining the methods that use those features (e.g., KDCoE~\cite{DBLP:conf/ijcai/ChenTCSZ18}). Especially \DOne\ is a dataset that has been very challenging for state-of-the-art methods~\cite{DBLP:journals/datamine/FanourakisEKC23,DBLP:journals/pvldb/SunZHWCAL20}.
    \item \textbf{\DThree} and \textbf{\DFour}~\cite{DBLP:conf/icml/GuoSH19} are the sparse and the dense versions, respectively, of another dataset that was constructed from DBpedia and Wikidata KGs, using Segment-based Random PageRank Sampling (SRPRS), a sampling method that tries to better capture real-world entity distributions (i.e., the number of triples in which an entity is involved). 
    \item \textbf{\DFive}~\cite{DBLP:conf/bigdataconf/EfthymiouSC15} is a dataset consisting of BBC music\footnote{\url{http://csd.uoc.gr/~vefthym/minoanER/datasets/bbcMusic.tar.gz}} and the BTC2012 version of DBpedia\footnote{\url{http://km.aifb.kit.edu/projects/btc-2012/}}. It contains various entity types such as musicians, their birth places and bands, while it also has the highest number of average attributes per entity. This dataset is well suited for EA methods that perform schema alignment, since its two KGs follow similar schema naming conventions.
    \item \textbf{\DSix, \DSeven} and \textbf{ \DEight}~\cite{DBLP:conf/ijcai/SunHZQ18} are three multilingual datasets, extracted from the English (EN), French (FR), Japanese (JA) and Chinese (ZH) versions of DBpedia\footnote{\url{http://downloads.dbpedia.org/2016-04/}}, describing actors, musicians, films, songs, writers, cities, football players and teams. 
    \item \textbf{\DNine}\ and \textbf{\DTen} are two EA datasets from the Integrated Crisis Early Warning System (ICEWS) and the two general KGs: Wikidata and Yago. ICEWS is a representative domain-specific KG, which contains political events with time annotations that embody the temporal interactions between politically related entities. Wikidata and Yago are two common KGs with extensive general knowledge that can provide background information. In both datasets, KGs exhibit low overlap compared to the previous datasets, meaning that they do not conform to the one-to-one assumption as most of the previous datasets do, while also the structural heterogeneity is expected high. In addition, these datasets do not come with attribute triples, so for in our experiments we used the entity names and the property ``has\_name'' for constructing some attribute triples. Finally, a key difference of these and the previous datasets is that the later come with some timestamps that are not exploited by the methods we experimentally evaluated.
    
\end{itemize}
Table~\ref{tab:datasets} details the following characteristics of datasets: 
\begin{itemize}
    \item Number of entities $|E|$: the number of nodes in each KG, each describing a real-world entity.
    \item Number of relation triples $|X|$: the number of triples whose head and tail (subject and object) are both entities.
    \item Number of attribute triples $|Y|$: the number of triples whose head (subject) is an entity and tail (object) is a literal value. 
    \item Number of relations $|R|$: the number of distinct relations appearing in a KG, i.e., the number of relation types.
    \item Number of attributes $|A|$: the number of distinct attributes appearing in a KG, i.e., the number of attribute types.
    \item Largest weakly connected component $maxCS$: defined in~\cite{DBLP:conf/esws/FanourakisECKPS23} as $\max_{CC \in wcc}(|CC|) / |E|$, where $wcc$ is the set of weakly connected components in a KG. Intuitively, the higher the size of the largest component (measured in number of nodes in that component), the easier it is to perform EA. 
    \item Weakly connected components ratio $wccR$: defined in~\cite{DBLP:conf/esws/FanourakisECKPS23} as $|wcc|/|E|$, where $|wcc|$ is the number of weakly connected components in a KG. Intuitively, for a fixed $|E|$, the higher the number of connected components, the more difficult it is to perform EA.
    \item Nodes without a name: the number of nodes for which no attribute can be used as an entity label. The set of attributes that can be used for that purpose (e.g., rdfs:label, skos:name, foaf:name) may be selected manually per KG, or detected automatically by statistical methods~\cite{DBLP:conf/edbt/Efthymiou0SC19}. 
    \item Data splitting in train/validation/test: the number of positively labeled (i.e., matching) seed alignment pairs used for the training, validation, and test set, respectively.
\end{itemize}
\vspace{-.5cm}
\subsection{Heterogeneity metrics}
\label{sec:heterogeneity_metric}
In this section, we propose different metrics that could enrich our understanding of the levels of heterogeneity the datasets exhibit. To this order, we propose 4 metrics: 2 structural metrics and 2 semantic metrics. Each of them is computed by only taking into account the aligned pairs, with the idea that structurally and/or semantically similar entities are more likely to be matched than dissimilar ones. Thus, the difficulties come when the entities that have to be matched are structurally and/or semantically dissimilar. 
Table \ref{tab:datasets_stat} reports for each dataset the following metrics: 
\begin{table*} [t]
\centering
\caption{Heterogeneity metric per dataset. See Section \ref{sec:heterogeneity_metric} for information about the different computed metrics.}
\label{tab:datasets_stat}
\begin{tabular}{l|cccccc} 
\textbf{Dataset}            
& Jaccard
& LDMAD
& Mean degree $KG_1$
& Mean degree $KG_2$
& Levenshtein on attributes
& Levenshtein on names \\
\hline
\textbf{\DOne} & 0.63 & 2.08 & 4.21 & 5.14 & 0.58 & 0.50 \\ 
\hline
\textbf{\DTwo} & 0.67 & 3.22 & 8.14 & 9.87 & 0.64 & 0.55 \\ 
\hline
\textbf{\DThree} & 0.62 & 1.51 &4.38 & 4.27 & 0.72 & 0.50 \\
\hline
\textbf{\DFour} & 0.68 & 2.13 & 7.29 & 7.81 & 0.73 & 0.59\\
\hline
\textbf{\DFive} & 0.67 & 1.33 & 1.79 & 2.93 & 0.72 & 0.71 \\
\hline

\textbf{\DSix} & 0.41 & 4.11 & 10.63 & 11.24 & 0.72 & 0.82 \\
\hline

\textbf{\DSeven} & 0.41 & 3.28 & 7.51 & 8.95 & 0.63 & 0.17 \\ 
\hline

\textbf{\DEight} & 0.41 & 3.67 & 7.04 & 9.20 & 0.67 & 0.02\\ 
\hline

\textbf{\DNine} & 0.02 & 90.21 & 102.19 & 22.45 & 0.85 & 0.92\\
\hline

\textbf{\DTen} & 0.03 & 43.77 & 47.76 & 5.94 & 0.41 & 0.95\\
\hline
\end{tabular}
\end{table*}


\begin{itemize}
    \item \textbf{Jaccard index between the matched neighbors (J): } Intuitively, matching entities whose neighbors are also respectively matched is easier than matching entities for which none of their neighbors are respectively matched. We thus define the set of entities in the neighborhood of $u$ that are matched with an entity in the neighborhood in $v$ as $M_{u,v} = \{(e_i, e_j) \in E_1 \times E_2 \mid e_i \equiv e_j, e_i \in \mathscr{N}(u), e_j \in \mathscr{N}(v)\}$ where $u$ and $v$ are two aligned entities and $\mathscr{N}(u)$ and $\mathscr{N}(v)$ their respective neighborhood. Our Jaccard index can then be defined as 
    $$J = \frac{1}{|M|}\sum\limits_{(u,v) \in M} \frac{|M_{u,v}|}{|\mathscr{N}(u)|+|\mathscr{N}(v)|-| M_{u,v}|},$$
    where $M$ is the set of matching pairs.
    
    \item \textbf{Local Degree Mean Absolute Deviation (LDMAD)}: The Jaccard index might be too restrictive in some cases, e.g., outside the one-to-one constraint, where matching entities might be far from each other in the same graph, leading to a poor Jaccard index. We thus compare the degree difference between two matched entities:  
    $$\text{LDMAD} = \frac{1}{|M|}\sum\limits_{(u,v) \in M} ||\mathscr{N}(u)| - |\mathscr{N}(v)||.$$
    In order to have a more precise idea on the impact of this metric, we also compute the mean degree of each graph.
    
    \item \textbf{Levenshtein Similarity of Attributes and Names (lev)}: Concerning the semantic similarity, we computed the Levenshtein similarity independently on both attribute values $a$ and entity names $b$: 
    $$\text{lev}_{\text{index}}(a,b) = \frac{|a| + |b| - \text{lev}_{\text{distance}}(a,b)}{|a|+|b|}.$$
    The name similarity is pretty straightforward for evaluating how semantically similar matched entities are. However, it might not be sufficient enough, e.g. in the case of multi-lingual datasets, where the names will be necessarily different. We thus also compute the Levenshtein similarity between the attributes of matching entities, first because their diversity could allow to bypass the language barrier and alphabet differences, e.g., with dates, and second because they are also one of the main focus of our semantic component. 
\end{itemize}

\subsection{Datasets Analysis}\label{ssec:dataset_analysis}

In terms of structural heterogeneity, we observe three clusters according to the different metrics we computed, in particular our Jaccard index. With this metric, the monolingual datasets (\DOne,\ \DTwo,\ \DThree,\ \DFour\ and \DFive) stand out with a low structural heterogeneity, as shown by a Jaccard index between 0.63 and 0.68, which means that 2/3 of each aligned entity's neighbors are also matched. In terms of Local Degree Mean Absolute Deviation in the mono-lingual cluster, we first can see the impact of the Dense datasets (\DTwo\ and \DFour) compared to the Sparse datasets (\DOne\ and \DThree). In both cases, despite increasing largely the mean degree of both graphs, the Local Degree Deviation does not increase in the same proportion, which might indicate a lower structural heterogeneity for the Dense datasets. Concerning \DFive, the low value of LDMAD has to be compared with the low value of mean degree in both graphs, and in particular $KG_1$. This makes \DFive\ very particular, with very few neighbors per entity, which is supported by the low number of relation triples in this dataset, as it can be seen in Table~\ref{tab:datasets}. In conclusion, the mono-lingual cluster can be subdivided into 3 subsets, with the Dense datasets being the least structurally heterogenous, followed by the Sparse datasets and finally \DFive\ in its own subset. 

In terms of semantic heterogeneity, \DFive\ stands out once again among the monolingual datasets by exhibiting the highest Levenshtein similarity, in particular on the names, showing a particularly low semantic heterogeneity on this dataset. It must also be compared with the number of node without names, almost none in the case of \DFive, while significant for the other monolingual datasets. In conclusion, among the monolingual datasets, \DFive\ is really particular, showing a very low mean degree on both graphs, which might bias its structural heterogeneity, while also showing a low semantic heterogeneity.

The multilingual datasets are then forming the second cluster, with a medium structural heterogeneity and a Jaccard index equal to 0.41 for all of them. \DSix\ has a high deviation in number of nodes but also a high degree on both graphs on average, which once taken into account show a lower structural heterogeneity compared to the other multilingual datasets. On the semantic heterogeneity, the three of them are really different on the name level, with \DSix\ having among the lowest semantic heterogeneity while \DSeven\ and \DEight\ exhibit the highest semantic heterogeneity among all datasets. This can first be explained by the alphabet difference, which is not taken into account by the Levenshtein distance, and secondly by the high number of unnamed entities in both \DSeven\ and \DEight. Multilingual datasets finally exhibit an average Levenshtein similarity of attribute values.

The last cluster is formed by the two datasets, \DNine\ and \DTen, which drop the one-to-one constraint and exhibit the most important structural heterogeneity. This can be explained by the low overlap ratio between each dataset \cite{jiang2024toward} and the high density of each graph, as shown by the mean degree of both graphs, but this also means that the aligned entities are not localized in one section of each graph, but more likely distilled in the whole graph. However, those graphs present the lowest semantic heterogeneity on the name level, which seems to indicate that the aligned entities share very similar names. The two datasets diverge for the Levenshtein similarity on the attribute level, which seems to indicate \DTen\ has a lower global semantic heterogeneity than \DNine, and thus might be considered a more challenging dataset.



It might however be important to notice that, whether the datasets exhibit a strong structural or semantic heterogeneity, being able to consider both is mandatory. First in our analysis, no dataset appeared as purely structural or purely semantic (even though \DNine\ and \DTen\ could arguably be close to such definition), it is thus necessary to consider them both. Second, our analysis does not consider the variability with a graph. Indeed, some graphs may for example exhibit a low semantic heterogeneity in general, but still have some local parts that are highly semantically heterogeneous. For those two reasons, it is thus important to consider both heterogeneities. 

To conclude this section, we observe a high diversity within the characteristics of each dataset. Apart from \DFive, the monolingual datasets seem to exhibit a lower structural heterogeneity and a higher semantic heterogeneity than the other datasets, which could indicate that structure-based approaches might perform better on those datasets. On the other side, the very high structural heterogeneity of \DNine\ and \DTen\ with a very low semantic heterogeneity indicate that methods that do not rely exclusively on the structure should perform better than structure-based approaches. Finally, multilingual datasets provide neither high structural heterogeneity nor high semantic heterogeneity, which might explain their popularity in the experimental evaluation of embedding-based EA methods. Table~\ref{tab:dataset_summary} summarizes the different levels of heterogenity per dataset. 

\begin{table}[t]
    \centering
    \caption{Summary of the different types and levels of heterogeneity per dataset.}
    \begin{tabular}{l|c|c}
        \textbf{Dataset} & Semantic heterogen. & Structural heterogen.\\
        \hline
        \textbf{\DOne} & High & Low \\
        \textbf{\DTwo} & High & Low \\
        \textbf{\DThree} & High & Low \\
        \textbf{\DFour} & High & Low \\
        \textbf{\DFive} & Low & Low \\
        \textbf{\DSix} & Low & Medium\\
        \textbf{\DSeven} & Medium & Medium \\
        \textbf{\DEight} & Medium & Medium \\
        \textbf{\DNine} & Low & High \\
        \textbf{\DTen} & Low & High
    \end{tabular}
    \label{tab:dataset_summary}
\end{table}

\color{black}
\section{Experimental Evaluation}\label{sec:experiments}
In this section, we present the experimental setting, 
baselines, measures, and we analyze the experimental results, addressing effectiveness, an ablation study, and computational cost. We conclude this section with insights and lessons learned.

\subsection{Experimental Setting}\label{ssec:settings}

All experiments were performed on a server with 16 AMD EPYC 7232P @ 3.1 GHz cores, 64 GB RAM, one RTX-4090 GPU (24 GB) and Ubuntu 18.04.5 LTS. 

For the structural model, when running \systR, we set the hyperparameters as proposed by RREA~\cite{DBLP:conf/cikm/MaoWXWL20}: the embedding dimension is 100, the architecture consists of 2 layers with a single attention head and a dropout rate of 0.3, and the training process uses a learning rate of 0.005 with the RMSprop optimizer.

For the experiments with \systK, we use the hyperparameters suggested by Knowformer~\cite{10092525}. 
We use 12 layers with 4 self-attention heads and we employ dropout for all layers, including the input dropout (0.5), attention score dropout (0.1), feed-forward dropout (0.3), and output dropout (0.1). Residual is 0.5, addition loss is 0.1, soft label is 0.25, learning rate is 0.0005 and batch size is 2,048.

For the factual model, the hyperparameters setting is the following: $\alpha$ is 0.8, $\lambda$ is 3, learning rate is 0.00005, training batch size is 24, while $d_k$ is 768. As for the layers, we use one attention layer for each KG with attention size equal to 768 and input dimension varying depending on the number of attribute types of the KG. 
We use the truncated negative sampling method of BootEA~\cite{DBLP:conf/ijcai/SunHZQ18}, 
with 
two negative samples for every positive sample.

For both models, we use CSLS~\cite{DBLP:conf/iclr/LampleCRDJ18} equal to 2 for normalizing the similarity score of the entities based on the density of their 2 nearest neighbors in the embedding space. 
The maximum number of epochs for the factual model and the structural model when using \systK\ is 200, and 1200 when using \systR, the minimum is 10 for the structural and 5 for the factual model, while the training stops early if the model is measured 3 times with H@1 score lower than its best achieved so far. Early stopping for the semi-supervision cycles is also applied, with max cycles = 4. Finally, the seed alignment is split into 20\% training, 10\% validation, and 70\% test.

\subsection{Baselines}\label{ssec:baselines}
To evaluate the benefits of our method, we compare its performance against eleven state-of-the-art baseline methods, namely
MTransE~\cite{DBLP:conf/ijcai/ChenTYZ17},
Knowformer~\cite{10092525},
BERT-INT~\cite{DBLP:conf/ijcai/Tang0C00L20}, RREA~\cite{DBLP:conf/cikm/MaoWXWL20},  COTSAE~\cite{DBLP:conf/aaai/YangLZWX20}, PipEA~\cite{DBLP:journals/corr/abs-2402-03025}, ZeroEA~\cite{DBLP:journals/pvldb/HuoCKNHLLNLQ24}, SelfKG~\cite{DBLP:conf/www/LiuHWCKD022}, Simple-HHEA~\cite{DBLP:conf/www/JiangXSWSSSLGS24}, AttrGNN~\cite{DBLP:journals/corr/abs-2010-03249} and PARIS+~\cite{DBLP:journals/pvldb/LeoneHAGW22}. 
For Knowformer, ZeroEA, SelfKG, Attr-GNN and Simple-HHEA we used directly the source code provided by the authors~\cite{git_knowformer,git_zeroea,git_selfkg,git_attr_gnn,git_simple_hhea}, while for MTransE, BERT-INT, RREA and PARIS+ we used the source code~\cite{git_study} provided by~\cite{DBLP:journals/datamine/FanourakisEKC23}. 
We were not able to reproduce the results of COTSAE~\cite{DBLP:conf/aaai/YangLZWX20} from the provided github link~\cite{git_cotsa}, so we report the results presented in~\cite{DBLP:conf/aaai/YangLZWX20} for the shared datasets.
For PipEA, the provided source code~\cite{git_pipea} only contains a non-iterative (i.e., basic) version. 

\vspace{-.5cm}
\subsection{Evaluation measures}\label{ssec:measures}
Traditionally, entity matching methods have been evaluated using classification metrics, like precision and recall, comparing the predicted matches to the ground-truth matches~\cite{DBLP:books/daglib/0030287,DBLP:series/synthesis/2015Christophides}. On the contrary, KG embedding-based EA methods, adopted rank-based evaluation metrics (e.g., hits at $k$). 
This is probably the result of the strong one-to-one assumption, considering that every entity in one KG should be matched with one entity in the other KG. We acknowledge that the criticism of~\cite{DBLP:journals/pvldb/LeoneHAGW22} is valid, i.e., that this assumption is unrealistic, and we further claim that under the one-to-one assumption, there is no need to perform entity alignment (knowing that everything is match, implies knowing all the matches in advance).
Nonetheless, for evaluating our method on an equal basis with existing methods, and in order to re-use the available open-source code as much as possible, we adopt the latter evaluation metrics.

The measures used for the effectiveness evaluation are the standard ones used in the literature, i.e., hits at $k$ (H@$k$) and Mean Reciprocal Rank (MRR). In order to reuse the code of PARIS+~\cite{DBLP:journals/pvldb/LeoneHAGW22} (that reports only precision, recall and F1-score), in Table~\ref{tab:effectiveness} we consider the F1-score equal to H@1 since in the test phase, each source entity gets a list of candidates~\cite{DBLP:journals/pvldb/SunZHWCAL20}. Furthermore, we also evaluate the effectiveness of the reciprocity filter alone, which is used for enriching the training set. As this filter returns final matching decisions, and not rankings of pairs, we evaluate its effectiveness using the standard classification measures, i.e., precision, recall, and F1-score, cumulatively, across all cycles, per dataset and per component (structural and factual). For example, assuming two execution cycles and a test set of 100 matching pairs, if the structural component returns 10 matches in the first cycle and 5 matches in the second cycle, all of which are correct, its cumulative precision is 100\% and its cumulative recall is 15\%. 


\vspace{-.5cm}
\subsection{\syst\ Performance}\label{ssec:results}
In this section, we present and analyze our experimental findings. First, we present the overall effectiveness results in mono- and multilingual datasets, and then proceed with a detailed ablation study.

\subsubsection{Effectiveness Results}

\begin{figure} [t]
    \centering
    \includegraphics[width=0.50\textwidth]{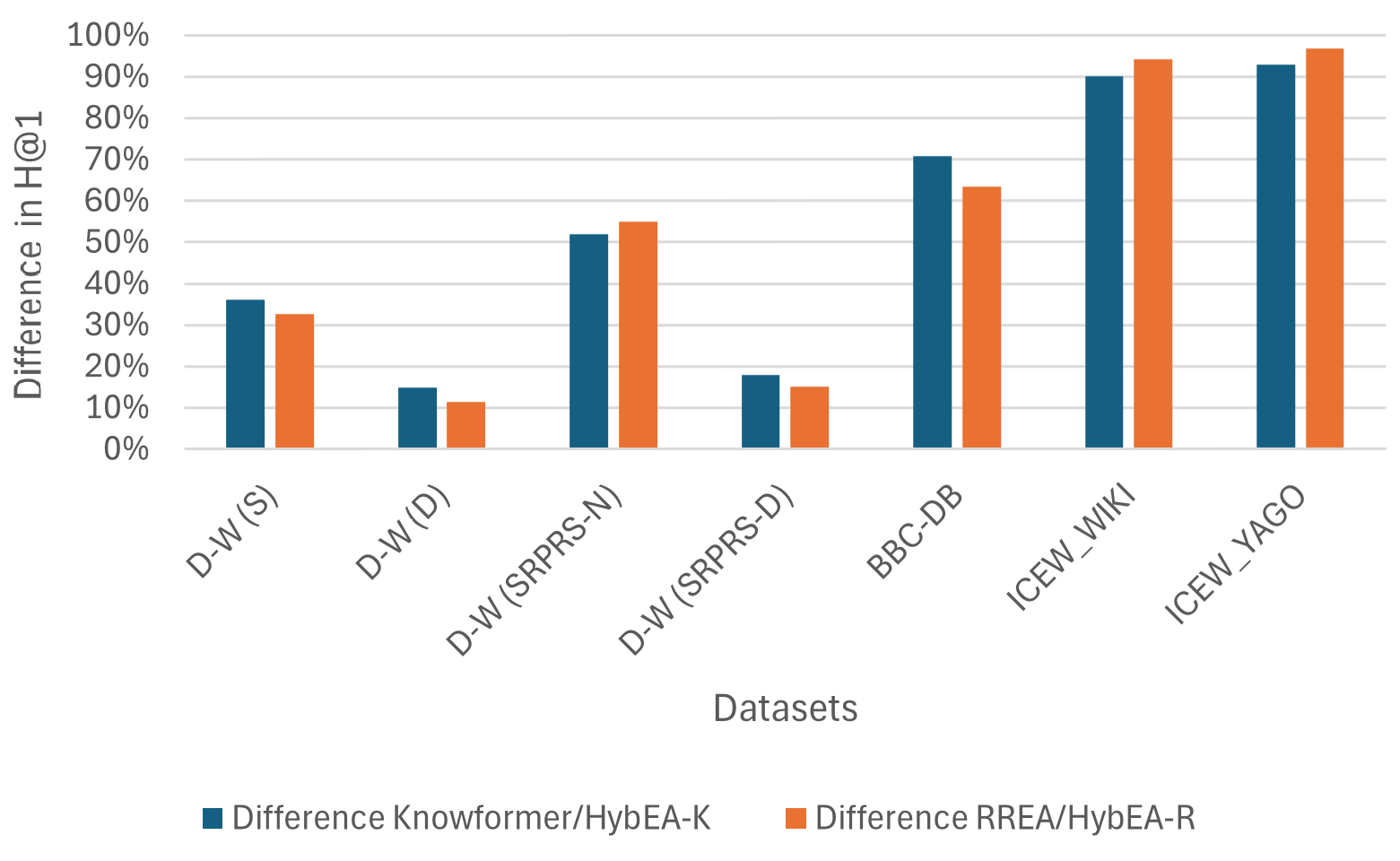}
    \caption{Difference in H@1 between \syst\ and the structural method employed.}
    \label{fig:performance_comparison}
\end{figure}

\textbf{Performance comparison between \syst\ and structural methods.} 
Figure~\ref{fig:performance_comparison} shows that no matter the dataset, \syst\ outperforms the structural method used alone.
In fact, when we compare our method using RREA as the structural component (i.e., \systR) and RREA basic, we achieve results that are from 11.35\% to 96.75\% better.  
A similar improvement is observed when comparing Knowformer to  \syst\ using Knowformer (i.e., \systK), we observe an improvement ranging from 14.8\% to 92.77\%, depending on the dataset.

Although Knowformer benefits greatly from the semi-supervised framework, it performs worse than RREA when used alone. Additionally, when included as the structural component, we observe that \systK \ is less efficient than \systR.

\begin{table*}[t]
    \centering
    \caption{Effectiveness results, with highest score per dataset and measure in boldface, runner-up underlined, best baseline (wrt. H@1) in red. The first iteration of semi-supervised methods reported separately, in italics. PARIS+: We report F1-score as H@1 (see Section~\ref{ssec:measures}). The last row ($\Delta$) reports the difference, in absolute number and in percentage, of the highest H@1 score (in bold) from the best baseline H@1 score (in red).}
    \label{tab:effectiveness}
    \begin{tabular}{ll|rrrrrrr}
        \textbf{Method} & \textbf{Metric} & \textbf{\DOne} & \textbf{\DTwo} & \textbf{\DThree} & \textbf{\DFour} & \textbf{\DFive} & \textbf{ICEWS-WIKI} & \textbf{ICEWS-YAGO} \\ \hline
        
        \multirow{3}{*}{\textbf{MTransE}} 
        & H@1 & 0.260 & 0.262 & 0.210 & 0.347 & 0.249 & 0.001 & 0.000\\ 
         & H@10  & 0.540 & 0.574 & 0.493 & 0.680 & 0.502 & 0.006 & 0.001\\
         & MRR & 0.35 & 0.36 & 0.30 & 0.46 & 0.33  & 0.004 & 0.001 \\ \hline

         \multirow{3}{*}{\textbf{Knowformer}} 
         & H@1 & 0.559 & 0.840 & 0.388 & 0.788 & 0.289 & 0.016 & 0.013\\ 
         & H@10  & 0.786 & 0.941 & 0.656 & 0.924 & 0.506 & 0.075 & 0.046\\
         & MRR & 0.64 & 0.87 & 0.47 & 0.83 & 0.37 & 0.03 & 0.02 \\ \hline
         
         \multirow{3}{*}{\textbf{BERT-INT}} 
         & H@1 & 0.440  & 0.426  & 0.519 & 0.642 & 0.925 & 0.561 & 0.756\\
         & H@10 & 0.489 & 0.485 & 0.534 & 0.650 & 0.937 & 0.700 & 0.859\\
         & MRR & 0.45 & 0.44 & 0.52 & 0.64 & 0.93 & 0.60 & 0.79 \\ \hline

          \multirow{3}{*}{\textbf{\textit{\shortstack[l]{RREA \\ (basic)}}}} 
         & H@1 & 0.655 &  0.878 & 0.446 & 0.817 &  0.436 & 0.050 & 0.026\\
         & H@10  & 0.884 & 0.986 & 0.754 & 0.962 & 0.652 & 0.227 &  0.136\\
         & MRR & 0.74 & 0.92 & 0.55 & 0.87 & 0.52 & 0.11 & 0.06 \\ \hline
        
         \multirow{3}{*}{\textbf{RREA}} 
         & H@1 & 0.718 & 0.937  & 0.503 & 0.881 & 0.468 & 0.050 &  0.027\\
         & H@10  & 0.900  & 0.991  & 0.768 & 0.977 & 0.651 & 0.230 & 0.142 \\
         & MRR & 0.79 & 0.96 & 0.59 & 0.91 & 0.54 & 0.11 & 0.06 \\ \hline
         
         \multirow{3}{*}{\textbf{COTSAE}}  & Hits@1 & - & - & \textcolor{red}{0.709} & \textcolor{red}{0.922}  & -  & -  & - \\
         & H@10 & - & - & 0.904 & 0.983  & -   & -  & -\\
         & MRR & - & - & 0.77 & 0.94 & -   & -  & -\\ \hline
         
         \multirow{3}{*}{\textbf{\textit{\shortstack[l]{PipEA \\ (basic)}}}} 
         & H@1 & 0.402 & 0.736 & 0.241 & 0.708 & 0.140 & N/A   & N/A\\
         & H@10 & 0.544 & 0.804 & 0.371 & 0.823 & 0.493 & N/A   & N/A\\
         & MRR & 0.45 & 0.76 & 0.29 & 0.75 & 0.02 & N/A   & N/A\\ \hline
         
        \multirow{3}{*}{\textbf{ZeroEA}} 
         & H@1 & 0.466 & 0.538 & 0.469 & 0.604 & \textcolor{red}{0.969} & N/A   & N/A \\
         & H@10 & 0.504 & 0.567 & 0.488 & 0.620 & 0.976 & N/A   & N/A \\
        & MRR & 0.43 & 0.48 & 0.45 & 0.55 & 0.80 & N/A   & N/A\\ \hline
         
        \multirow{3}{*}{\textbf{SelfKG}} 
         & H@1 & 0.542 & 0.620 & 0.586 & 0.734 & 0.961 & \textcolor{red}{0.839} & 0.806 \\
         & H@10 & 0.707 & 0.749 & 0.750 & 0.852 & 0.989 & 0.931 & 0.867\\
         & MRR & 0.59 & 0.66 & 0.63 & 0.77 & 0.97 & 0.871 & 0.828 \\ \hline
         
        
        \multirow{3}{*}{\textbf{Simple-HHEA}} 
        & H@1 & 0.071 & 0.077 & 0.104 & 0.144 & 0.098 & 0.720 & \textcolor{red}{0.847}\\ 
         & H@10  & 0.235 & 0.327 & 0.305 & 0.402 & 0.291 & 0.872 & 0.915\\
         & MRR & 0.12 & 0.15 & 0.17 & 0.22 & 0.16 & 0.754 & 0.870\\ \hline
         
        \multirow{3}{*}{\textbf{AttrGNN}} 
         & H@1 & 0.522 & 0.570 & 0.366 & 0.193 & 0.311 & 0.047 & 0.015 \\ 
         & H@10  & 0.692 & 0.657 & 0.588 & 0.343 & 0.539 & - & - \\
         & MRR & 0.58 & 0.60 & 0.44 & 0.24 & 0.39 & 0.09 & 0.04 \\ \hline
        
        \multirow{1}{*}{\textbf{PARIS+}} 
         & H@1 & \textcolor{red}{0.841} & \textcolor{red}{0.938} & 0.442 & 0.834 & 0.387 & 0.672 & 0.687\\
         \hline \hline
         
        \multirow{3}{*}{\textbf{\systR}}
         & H@1 & \textbf{0.989} & \textbf{1.000}  & \textbf{0.972}  & \textbf{0.997}  & \underline{0.993} & \textbf{0.994} & \textbf{0.994} \\
         & H@10  & \textbf{0.997} & \textbf{1.000} & \textbf{0.992} & \textbf{1.000}  & \textbf{1.000} & \textbf{0.997} & \textbf{0.997} \\
         & MRR & \textbf{0.99} & \textbf{1.00} & \textbf{0.98} & \textbf{1.00} & \textbf{1.00} & \textbf{1.00} & \textbf{1.00} \\ \hline
         
         \multirow{3}{*}{\textbf{\systK}} 
         & H@1 & \underline{0.920} & \underline{0.988}  & \underline{0.908}  & \underline{0.968}  & \textbf{0.996} & \underline{0.916} & \underline{0.941} \\
         & H@10  & \underline{0.969} & \underline{0.997} & \underline{0.954} & \underline{0.987}  & \underline{0.997} & \underline{0.938} & \underline{0.944}\\
         & MRR & \underline{0.93} & \underline{0.99} & \underline{0.92} & \underline{0.97} & \underline{0.99} & \underline{0.92} & \underline{0.94}\\ \hline \hline 
         
         
         \multirow{2}{*}{\textbf{$\Delta$}} & \multirow{2}{*}{H@1} & +0.148 & +0.062 & +0.263 & +0.075 & +0.027 & +0.155 & +0.147\\
         &  & (+17.6\%) & (+6.6\%) & (+40.6\%) & (+7.7\%) & (+3.6\%) & (+18.5\%) & (+17.4\%) \\

    \end{tabular}
\end{table*}
\begin{table}[t]
    \centering
    \caption{Effectiveness results in multilingual datasets, with highest score per dataset and measure in boldface; runner-up underlined. OOM: out of memory.}
    \label{tab:mul_effectiveness}
    \begin{tabular}{ll|rrrrr}
        \textbf{Method} & \textbf{Metric} & \textbf{\DSix} & \textbf{\DSeven} & \textbf{\DEight} \\ \hline
        
        \multirow{3}{*}{\textbf{MTransE}} 
         & H@1 & 0.244 & 0.279 & 0.308 \\
         & H@10 & 0.556 & 0.575 & 0.614  \\
         & MRR & 0.335 & 0.349 & 0.364  \\ \hline
        
         \multirow{3}{*}{\textbf{Knowformer}} 
         & H@1 & 0.774 & 0.731 & 0.765 \\
         & H@10  & 0.932 & 0.902 & 0.888 \\
         & MRR & 0.832 & 0.793 & 0.811 \\ \hline
         
        \multirow{3}{*}{\textbf{BERT-INT}} 
         & H@1 & 0.992 & 0.964 & 0.968 \\
         & H@10 & 0.998 & 0.991 & 0.990 \\
         & MRR & 0.995 & 0.975 & 0.977 \\ \hline
         
        \multirow{3}{*}{\textbf{RREA (basic)}} 
         & H@1 & 0.739 & 0.713 & 0.715 \\
         & H@10 & 0.946 & 0.933 & 0.929 \\
         & MRR & 0.816 & 0.793 & 0.794 \\ \hline
         
        \multirow{3}{*}{\textbf{RREA}} 
         & H@1 & 0.827 & 0.802 & 0.801 \\
         & H@10 & 0.966 & 0.952 & 0.948 \\
         & MRR & 0.881 & 0.858 & 0.857 \\ \hline
        
         
         
        \multirow{3}{*}{\textbf{ZeroEA}} 
         & H@1 & \underline{0.998} & \underline{0.982} & \underline{0.985} \\
         & H@10  & \underline{0.999} & \underline{0.995} & {0.993} \\
         & MRR & \underline{0.998} & \underline{0.989} & \underline{0.991} \\ \hline
         
        \multirow{3}{*}{\textbf{SelfKG}} 
         & H@1 & 0.957 & 0.813 & 0.742 \\
         & H@10  & 0.992 & 0.906 & 0.861 \\
         & MRR & 0.971 & 0.844 & 0.782 \\ \hline
         
        
        
        \multirow{3}{*}{\textbf{AttrGNN}} 
         & H@1 & 0.942 & 0.783 & 0.796\\
         & H@10 & 0.986 & 0.920 & 0.929\\
         & MRR & 0.959 & 0.834 & 0.845\\ \hline
        
        \multirow{1}{*}{\textbf{PARIS+}} 
         & H@1 & 0.882 & 0.824 & OOM \\
         \hline \hline
         
        \multirow{3}{*}{\textbf{\systR}} 
         & H@1 & \textbf{0.999} & \textbf{0.993} & \textbf{0.994} \\
         & H@10 & \textbf{1.000} & \textbf{0.999} & \textbf{0.999} \\
         & MRR & \textbf{0.999} & \textbf{0.995} & \textbf{0.996} \\ \hline
         
         
\end{tabular}
\end{table}

\textbf{Results analysis} As presented in Section \ref{sec:dataset_analysis}, the different datasets can be divided into 3 clusters: the monolingual datasets, which exhibit low structural heterogeneity and high semantic heterogeneity and in which \DFive\ appears as an outlier, the multilingual datasets, which neither show a particular structural nor semantic heterogeneity, and the 
datasets that drop the one-to-one assumption, which present a strong structural heterogeneity but which are semantically very similar.

Based on this clustering, we can draw some expectations on the behavior and the performances of each method. Structure-based approaches such as Knowformer, RREA, PipEA, MTransE, PARIS+ should perform better when the structural heterogeneity is low, i.e., in the monolingual datasets, and obtain very poor result on highly structurally heterogenous datasets, such as \DNine\ and \DTen. On the other side, we expect Language Model-based approaches such as BERT-INT, ZeroEA and SelfKG to perform better on semantically homogeneous datasets. Finally, most methods should be able to perform correctly on multilingual datasets, due to the low heterogeneity values on both structural and semantic levels. 

Results obtained on highly heterogeneous datasets both structurally and semantically are presented in Table~\ref{tab:effectiveness}, while results for the multilingual datasets that present lower heterogeneity are presented in Table~\ref{tab:mul_effectiveness}. In the last row of Table~\ref{tab:effectiveness}, we also report the difference ($\Delta$) of \syst\ to the best-performing baseline method, in terms of H@1, both in absolute numbers, as well as in a percentage improvement with respect to the best baseline H@1 score.

First of all, we can observe that most of our expectations have been met. On the datasets with a high semantic heterogeneity and a low structural heterogeneity, the structure-based approaches obtain better results. Among the baseline methods, RREA obtains the best results on both \DOne\ and \DTwo, with respectively 0.718 and 0.937 H@1. In comparison, \systR\ reaches 0.989 and 1.0 H@1, respectively, on these datasets, while \systK\ obtains 0.92 and 0.988 respectively, showing very significant improvement, in particular on \DOne. The same behavior can be observed on \DThree\ and \DFour, where the best score among the baseline methods are obtained by COTSAE with respectively 0.709 and 0.922 H@1, followed by RREA. On the other side, \systR\ obtains 0.972 and 0.997 H@1 on those datasets, showing an improvement of more than 40\% on \DThree.  Finally, \DFive\ appears as an outlier: methods that obtain best scores on this dataset are the Language Model-based approaches, such as ZeroEA which reaches 0.969 H@1 on this dataset, followed by SelfKG and BERT-INT, while Structure-based approaches obtain very low results on this dataset, for example Knowformer obtains 0.289 H@1 and RREA 0.468 H@1. We hypothesise that it might come from the particular structure of \DFive\ on one side, which shows a very low average degree and might show a graph made of multiple stars, where many nodes are only connected to one central node. This might lead to a difficult differentiation between the leaves of the star graphs by the Structure-based approaches. On the other side, the low value of semantic heterogeneity in particular on the entities names probably benefit to the Language Model-based approaches. \systR\ and \systK\ once again show very good scores with respectively 0.972 H@1 and 0.996 H@1, which seems to show the ability of \syst\ to take into account both structural and semantic information. However, due to the particularity of \DFive\ properties, the high score we observe could be the fact of another unseen property. In order to ensure it, we thus also compare to \DNine\ and \DTen.

In terms of structural heterogeneity, we thus also evaluate our approach on \DNine\ and \DTen, which show almost no structural similarities in our analysis on Section \ref{sec:dataset_analysis}. This is something that we can directly observe on the results presented by the Structure-based approaches, systematically close to 0 on these datasets. For example, RREA and Knowformer, which obtain among the best scores with the previous datasets, obtain a score of 0.050 H@1 and 0.016 H@1, respectively, on \DNine, and 0.027 H@1 and 0.013 H@1, respectively, on \DTen. On the other side, Language Model-based approaches obtain the best score, as for \DFive: even though we were not able to run ZeroEA on those dataset, SelfKG and BERT-int show good performance on these dataset, SelfKG obtaining the best score of 0.838 H@1 on \DNine. Finally, Simple-HHEA is a method that has been designed to perform on these two datasets, which explains its good performance here, where it obtains the best score on \DTen\ with 0.847 H@1, while obtaining very low results everywhere else. Once again, \systR\ and \systK\ present the best scores by improving the state-of-art by more than 17\% for each dataset, respectively, reaching 0.994 H@1 and 0.916 H@1 on \DNine\ and 0.994 H@1 and 0.941 H@1 on \DTen. These results confirm the ability of \syst\ to efficiently consider both structural and semantic information, in a very large range of dataset complexity, from datasets with a high semantic heterogeneity to datasets with a high structural heterogeneity, some of which even drop the one-to-one assumption.

It is also worth noticing that \systR\ obtains almost perfect accuracy on every dataset, reaching almost 1.0 H@1, regardless of its heterogeneity, showing that these datasets are now near to be solved. More importantly, we believe that the fact that \syst\ is able to use any structural method, and in particular here RREA and Knowformer, and is still able to obtain significantly better results then those methods alone is of significant importance. First, it shows that \syst\ allows to systematically improve the structural method it is based on. Second, it also shown that even though some datasets exhibit different kind of heterogeneities, being able to take into account every kind of information is mandatory. It is particularly important for \DNine\ and \DTen, whose levels of structural heterogeneity might indicate that the structural component is useless. Yet, the improvement in performances exhibited by \syst\ and in particular \systR\ compared to the baseline methods seems to contradict this point, showing that even with low levels of structural similarity, the structural component is able to improve the alignment. This point is discussed further in Section \ref{sec: ablation_study}.

Finally, in order to provide an analysis on a wide range of datasets, we conducted similar experiments on multilingual datasets, namely \DSix, \DSeven\ and \DEight. The results can be found in Table~\ref{tab:mul_effectiveness}. As shown in Section~\ref{sec:dataset_analysis}, these datasets do not exhibit a particular level of heterogeneity, and it can be observed by the results obtained by the different baselines. Both Structure-based and Language Model-based approaches manage to obtain significant results, with a clear advantage for Language Model-based approaches such as ZeroEA and BERT-INT, but Structure-based methods such as RREA and Knowformer also achieve honourable results. These results are on par with our previous dataset analysis. Once again, \systR\ achieves the best results on the three datasets, even though the differences are much less important due to the already high performance of the baselines.

\textbf{Reciprocity filter.} 
For a more detailed evaluation, we also present in Table~\ref{tab:reciprocity_rrea} the precision, recall and F1-score of the entity pairs added to the final results by the reciprocity filter of each component. We compute these measures cumulatively for the pairs suggested by each component over all cycles. For presentation purposes, we only show the results for \systR, and similar analysis for \systK\ can be found in Appendix \ref{Appendix_reciprocity}. Also, more detailed results per cycle show that the vast majority of pairs suggested by the reciprocity filter are provided in the first cycle, almost in all datasets (see Figure~\ref{fig:perf_cycle} in Appendix \ref{Appendix_robust}). Overall, the reciprocity filter is highly effective, yielding perfect precision and very high recall. Moreover, it remains computationally efficient and does not introduce significant processing time.

\begin{table*}[t]
    \centering
    \caption{Cumulative Precision (Pr), Recall (Re), and F1-score (F1) results of the reciprocity filter per dataset and per component, over all iterations. Separate evaluation for structural component ran first and for structural component ran first. Maximum recall (max) of the second component is 1 - Re of the first component.}
    \label{tab:reciprocity_rrea}
    \begin{tabular}{l|l|rrrr@{\hskip 1cm}|l|rrrr}
         \textbf{Dataset} & \textbf{structural first}  & \textbf{Pr} & \textbf{Re} & (max) & \textbf{F1} & \textbf{factual first} & \textbf{Pr} & \textbf{Re} & (max) & \textbf{F1} \\ \hline
         
         \multirow{3}{*}{\textbf{\DOne}} 
         & structural & 1.0 & 0.63 &  & 0.77 & fact.   & 1.0 & 0.54 &  & 0.70 \\ 
         & factual   & 1.0 & 0.23 & (0.37) & 0.37 & struct. & 1.0 & 0.36 & (0.46)  & 0.53 \\
         \cline{2-11}
         & \textbf{cumulative} & 1.0 & 0.86 &  & 0.92 & \textbf{cumulative} & 1.0 & 0.90 &  & 0.95 \\ 
         \hline
         \multirow{3}{*}{\textbf{\DTwo}} 
         & structural & 1.0 & 0.68 &  & 0.81 & fact.   & 1.0 & 0.60 &  & 0.75 \\ 
         & factual   & 1.0 & 0.22 & (0.32) & 0.36 & struct. & 1.0 & 0.38 & (0.40) & 0.55 \\ 
         \cline{2-11}
         & \textbf{cumulative} & 1.0 & 0.91 &  & 0.95 & \textbf{cumulative} & 1.0 & 0.98 &  & 0.99 \\ 
         \hline
         \multirow{3}{*}{\textbf{\DThree}} 
         & structural & 1.0 & 0.48 &  & 0.65 & fact.   & 1.0 & 0.52 &  & 0.69 \\ 
         & factual   & 1.0 & 0.33 & (0.52) & 0.50 & struct. & 1.0 & 0.33 & (0.48)  & 0.50 \\
         \cline{2-11}
         & \textbf{cumulative} & 1.0 & 0.82 &  & 0.90  & \textbf{cumulative} & 1.0 & 0.86 &  & 0.92 \\ 
         \hline
         \multirow{3}{*}{\textbf{\DFour}} 
         & structural & 1.0 & 0.71 &  & 0.83 & fact.   & 1.0 & 0.44 &  & 0.61 \\
         & factual   & 1.0 & 0.13 & (0.29) & 0.23 & struct. & 1.0 & 0.51 & (0.57) & 0.67 \\ 
         \cline{2-11}
         & \textbf{cumulative} & 1.0 & 0.85 &  & 0.91 & \textbf{cumulative} & 1.0 & 0.95 &  & 0.97 \\ 
         \hline
         \multirow{3}{*}{\textbf{\DFive}}
         & structural & 1.0 & 0.44 &  & 0.62 & fact.   & 1.0 & 0.73 &  & 0.85 \\ 
         & factual   & 1.0 & 0.49 & (0.56) & 0.65 & struct. & 1.0 & 0.22 &   (0.27)     & 0.36 \\
         \cline{2-11}
         & \textbf{cumulative} &1.0 & 0.94 &  & 0.96  & \textbf{cumulative} & 1.0 & 0.95 &  & 0.98\\ 
        \hline
        \multirow{3}{*}{\textbf{\DNine}}
         & structural & 1.0 & 0.02 &  & 0.04 & fact.   & 1.0 & 0.92 &  & 0.96 \\
         & factual  & 0.99 & 0.89 & (0.98) & 0.94 & struct. & 1.0 & 0.01 & (0.08) & 0.02 \\
         \cline{2-11}
         & \textbf{cumulative} & 0.99 & 0.91 &  & 0.95 & \textbf{cumulative} & 1.0 & 0.92 &  & 0.96 \\ 
         \hline
         \multirow{3}{*}{\textbf{\DTen}}
         & structural & 1.0 & 0.01 &  & 0.03 & fact.   & 1.0 & 0.94 &  & 0.97 \\
         & factual   & 1.0 & 0.92 & (0.99) & 0.96 & struct. & 1.0 & 0.003 & (0.06) & 0.02 \\
         \cline{2-11}
         & \textbf{cumulative} & 1.0 & 0.94 &  & 0.97 & \textbf{cumulative} & 1.0 & 0.94 &  & 0.97 \\ 
         \hline
    \end{tabular}
\end{table*}

It is essential to note that the order of the components is determining the results of Table~\ref{tab:reciprocity_rrea}, as the second component is only called for the pairs that the first component could not match with the reciprocity filter.
In other words, the highest achievable recall by the second component is 1 minus the recall of the first component, expressed by the (max) column. 

For further analyzing the previous note, the left half of the table shows the results obtained by running the structural component first and the right half of the table shows the results obtained by running the factual component first (i.e., the default setting). In both cases, we also report the maximum achievable recall (max) by the second component. By comparing the cumulative F1 scores for the same dataset in the two variations, we notice that by executing the factual component first, we achieve better results than executing the structural component first, in all datasets. We also observe that the contribution of the factual component, when executed second, is very limited, since the effectiveness of the structural component does not allow much room for improvement. On the other hand, when executing the factual component first, the contribution of the two components is more balanced. Note that those results only reflect the effectiveness of the reciprocity filter and they do not reflect the overall evaluation of \systR, although the reciprocity filter largely impacts the overall effectiveness of \systR.

Another very important observation comes up from the comparison of the H@1 scores of \systR\ and \systR\ (struct. first) in Table~\ref{tab:ablation_RREA} and the Recall results (since Precision is almost always 1.0) shown in the factual first and structural first parts of Table~\ref{tab:reciprocity_rrea}, respectively. We notice that the vast majority of matches identified by \systR\ come from the reciprocity filter and only a very small number of matches are left to be decided by the ranking of the most likely matches returned by the last component. 

In more detail, we observe the following: 
\begin{itemize}
    \item \DOne: When executing the structural component first, the cumulative recall from the reciprocity filter is 0.86 and H@1 of \systR\ (struct. first) is 0.87, i.e., a 1\% difference. 
    When executing the factual component first, the cumulative recall from the reciprocity filter is 0.90, but H@1 of \systR\ is 0.99, i.e., a 9\% total difference.

    \item \DTwo: When executing the structural component first, the cumulative recall from the reciprocity filter is 0.91 and H@1 of \systR\ (struct. first) is also 0.91, i.e., the number of correct matches identified outside the reciprocity filter is negligible. 
    When executing the factual component first, the cumulative recall from the reciprocity filter is 0.98, but H@1 of \syst\ is 1.0, i.e., 2\% higher.

    \item \DThree: When executing the structural component first, the cumulative recall from the reciprocity filter is 0.82 and H@1 of \systR\ (struct. first) is also 0.82, showing again a negligible contribution of non-reciprocity-filter suggested matches. 
    When executing the factual component first, the cumulative recall from the reciprocity filter is 0.86, but H@1 of \systR\ is 0.97, i.e., an 11\% increase.

    \item \DFour: With the structural component executed first, the cumulative recall from the reciprocity filter is 0.85 and H@1 of \systR\ (struct. first) is 0.85; once again a negligible contribution of the matches not provided by the reciprocity filter. 
    When executing the factual component first, the cumulative recall from the reciprocity filter is 0.95, but H@1 of \systR\ is 0.99, i.e., a 4\% increase.

    \item \DFive: The cumulative recall from the reciprocity filter when executing the structural component first is 0.94, while H@1 of \systR\ (struct. first) is 0.94, once again showing no difference.
    When executing the factual component first, the cumulative recall from the reciprocity filter is 0.95, but H@1 of \systR\ is 0.99, i.e., an almost 4\% total difference.
    
    \item \DNine: The cumulative recall obtained by the reciprocity filter when using the structural component first is 0.91, and H@1 of \systR\ (struct. first) is 0.98. This is the first time we observe a significant difference between the recall of the reciprocity filter and the performance of the model when using the structural component first.
    When executing the factual component first, the cumulative recall from the reciprocity filter is 0.92, but H@1 of \systR\ is 0.99, i.e., a 7\% total difference.
    
    \item \DTen: The cumulative recall obtained by the reciprocity filter when using the structural component first is 0.94, and H@1 of \systR\ (struct. first) is 0.95, i.e., a 1\% difference.
    When executing the factual component first, the cumulative recall from the reciprocity filter is 0.94, but H@1 of \systR\ is 0.99, i.e., a 7\% total difference.
\end{itemize}

From the above findings, we can observe that the effectiveness of the reciprocity filter is on par with the results exhibited by \systR: both are higher when using the factual component first.
It is also interesting to note that the highest differences observed between the cumulative recall of the reciprocity filter and H@1 are on \DOne, \DThree\ and \DNine, which are the datasets that showed the most important heterogeneity in the dataset analysis section, \DOne\ and \DThree\ for the semantic and \DNine\ for the structural. It can thus be explained by the fact that one of the component has more difficulties to correctly align, leading to a significant difference between the reciprocity filter and the performance.

\textbf{Adaptability of \syst.} 
In this section, we examine the contribution of both structural and factual components of \syst, keeping in mind that in each cycle, the factual component is executed first, followed by the structural component. As we can observe from Table~\ref{tab:reciprocity_rrea}, a higher contribution of the structural component, compared to the factual component, is only observed in dense datasets (0.51 recall in \DFour). On the contrary, a higher contribution of the factual component is observed, when the structural heterogeneity is high (0.54 recall in \DOne, 0.60 in \DTwo, 0.52 in \DThree, and 0.73 in \DFive). The extreme cases are \DNine\ and \DTen, where regardless of the order of executing the two components, only the factual component contributes to the results.

When reading the table row by row, it is interesting to observe that, although in many datasets, e.g., in \DOne,  \DTwo, \DFour, the contribution of the structural component, when ran first (left part), is greater than the contribution of the factual component, when ran first (right part), yet, in all datasets, the cumulative scores when runnning the factual component first are better. This confirms that when we run the factual component first, we better exploit both models, i.e., the second models can better cover the missed matches of the first model. Concluding, the exploitation of both structural and factual information by our approach, using two different models, leads to higher adaptability of our method, enhancing the overall performance.


\begin{figure}
    \centering
    \includegraphics[width=\linewidth]{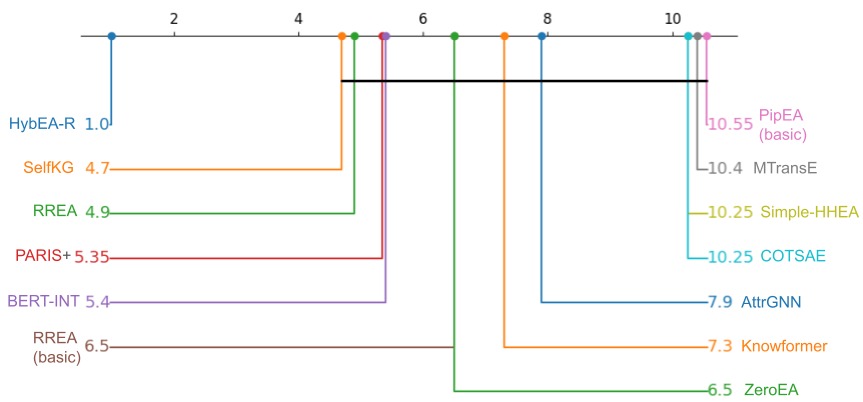}
    \caption{Statistical significance rankings of EA methods according to H@1.}
    \label{fig:bonferoni}
\end{figure}

\textbf{Effectiveness ranking of EA methods.} 
\label{sec:ranking}
To infer a statistically significant ranking of the methods, we rely on the non-parametric Friedman test~\cite{DBLP:journals/jmlr/Demsar06}. The null hypothesis $H_0$ of the Friedman test is that ``The mean performance for each method is equal'', while the alternative hypothesis $(H_a)$ states exactly the opposite. With p-value $2.84 × 10^{-9}$ of Friedman test for Hits@1, we can reject the null hypothesis $H_0$ at a $5\%$ confidence-level $(\alpha)$. In the sequel, we conduct the Bonferoni post-hoc test to compare the methods pairwise. This test reports as significance the average ranks of two methods if they differ by a critical distance (CD) given by $q_{\alpha} \sqrt{\frac{k(k+1)}{6 N}}$, where $N$ is the number of the datasets, $q_{\alpha}$ is a constant based on $\alpha$, and $k$ is the number of methods in total. For 13 EA methods, 10 datasets and $\alpha=0.05$, the value of $CD$ is 3.93.

From Figure~\ref{fig:bonferoni}, we can see that \systR~ outperforms all the other EA methods with statistically significant difference, since the rankings of the other methods have a distance from the ranking of \systR~ greater than the CD. SelfKG and then RREA has the closest ranking to \systR, but still with higher than the CD distance. MTransE and PipEA(basic) have the highest ranking, meaning that they are the most distant methods.

\begin{table*}[t]
    \centering
    \caption{Ablation study for \systR.}
    \label{tab:ablation_RREA}
    \begin{tabular}{ll|rrrrrrr}
        \textbf{Method} & \textbf{Metric} & \textbf{\DOne} & \textbf{\DTwo} & \textbf{\DThree} & \textbf{\DFour} & \textbf{\DFive} & \textbf{ICEWS-WIKI} & \textbf{ICEWS-YAGO} \\ \hline

        \multirow{3}{*}{\textbf{\systR}}
         & H@1 & \textbf{0.989} & \textbf{1.000}  & \textbf{0.972}  & \textbf{0.997}  & \textbf{0.993} & \textbf{0.994} & \textbf{0.994} \\
         & H@10  & \textbf{0.997} & \textbf{1.000} & \textbf{0.992} & \textbf{1.000}  & \textbf{1.000} & \textbf{0.997} & \textbf{0.997} \\
         & MRR & \textbf{0.99} & \textbf{1.00} & \textbf{0.98} & \textbf{1.00} & \textbf{1.00} & \textbf{1.00} & \textbf{1.00} \\ \hline

        \multirow{3}{*}{\textbf{\shortstack[l]{\systR \\ (basic)}}}
         & H@1 & 0.986 &  0.999 & 0.963  &  0.995 & 0.991 & \textbf{0.994} & \textbf{0.994} \\
         & H@10  & \textbf{0.997} & \textbf{1.000} & 0.989 & \textbf{1.000}  & \textbf{1.000} & \textbf{0.997} & \textbf{0.997} \\
         & MRR & \textbf{0.99} & \textbf{1.00} & 0.97 & \textbf{1.00} & 0.99 & \textbf{1.00} & \textbf{1.00} \\ \hline
         
        \multirow{3}{*}{\textbf{\shortstack[l]{\systR \\ (basic; str.first)}}}
         & H@1 & 0.779 & 0.866  &  0.742 &  0.797 & 0.832 & 0.985 & 0.953\\
         & H@10  & 0.802 & 0.883 & 0.781 &  0.809 & 0.843 & 0.995 & 0.976\\
         & MRR & 0.78 & 0.87 & 0.75 & 0.80 & 0.83 & 0.98 & 0.96\\ \hline
         
        \multirow{3}{*}{\textbf{\shortstack[l]{\systR \\ (struct. first)}}}
         & H@1 & 0.872 & 0.915  & 0.829  & 0.853  & 0.942 & 0.985 & 0.953\\
         & H@10  & 0.881 & 0.922 & 0.847 & 0.860  & 0.944 & 0.995 & 0.976\\
         & MRR & 0.87 & 0.91 & 0.83 & 0.85 & 0.94 & 0.98 & 0.96 \\ \hline

        \multirow{3}{*}{\textbf{\shortstack[l]{\systR \\ (w/o fact.)}}}
         & H@1 & 0.655 &  0.878 & 0.446 & 0.817 &  0.436 & 0.050 & 0.026\\
         & H@10  & 0.884 & 0.986 & 0.754 & 0.962 & 0.652 & 0.227 &  0.136\\
         & MRR & 0.74 & 0.92 & 0.55 & 0.87 & 0.52 & 0.11 & 0.06 \\ \hline
         
         \multirow{3}{*}{\textbf{\shortstack[l]{\systR \\ (w/o struct.)}}} 
         & H@1 & 0.558 & 0.621  & 0.513  & 0.444 & 0.748 & 0.926 & 0.948\\
         & H@10  & 0.612 & 0.672 & 0.592 & 0.488 & 0.769 & 0.974 & 0.975\\
         & MRR & 0.79 & 0.82 & 0.70 & 0.70 & 0.91 & 0.94 & 0.96\\ \hline

    \end{tabular}
\end{table*}

\subsubsection{Ablation Study.}\label{sec: ablation_study} 
We conduct an ablation study, showing that by changing any single component of \systR, the entire performance drops (Table~\ref{tab:ablation_RREA}). This demonstrates that all of our architectural choices have a positive impact on the effectiveness of \systR. A similar analysis for \systK, showing even bigger impact for each component, can be found in the Appendix (Table~\ref{tab:ablation_knowformer}).

\textbf{Impact of component ordering.} We experiment with different orders of execution for the structural and factual components, as presented in row \systR\ (struct. first) of Table~\ref{tab:ablation_RREA}. We observe that we get quite different results, with a clear winner in all datasets being running the factual component first (\systR). Consequently, we propose running the factual component first, as our default configuration.

\textbf{Impact of the semi-supervision step.} Table~\ref{tab:ablation_RREA} reports the results of our method at the end of the first cycle, when running the structural component first, denoted as \systR\ (basic; str.first), as well as when running the factual component first, denoted as \systR\ (basic). Compared to the results of the full pipeline, with the corresponding order of components, the basic version always yields lower or equivalent results. The difference between the basic version of \systR\ and the full version of \systR is not big (e.g., the difference between between \systK\ (basic) and \systK, reported in Table~\ref{tab:ablation_knowformer} of Appendix is much bigger), because \systR\ (basic) is already achieving almost perfect scores, so there is not much room for improvement. The difference between the basic and the full version is clearer when comparing the results of \systR\ (basic; str. first) with \systR\ (struct. first). 
It is important to note here that the basic version of \systR\ already outperforms all the state-of-the-art baseline methods. 

\textbf{Impact of the factual component.} 
The row \systR\ (w/o fact.) skips the factual component and essentially runs the structural component of \systR\ only (i.e., RREA), enriched with the reciprocity filter.  This experiment involves only one cycle, since re-training the same model without new pairs suggested from the other model, yields the same results in subsequent cycles. Since the reciprocity filter results show perfect precision (in Table~\ref{tab:reciprocity_rrea}), the results of \systR\ (w/o fact.) in Table~\ref{tab:ablation_RREA} and the results of RREA (basic) in Table~\ref{tab:effectiveness} are identical in all datasets.

We observe that \emph{the factual component has a huge impact} on the results, since, when we remove it, the H@1 scores drop by 12.2\% to even 96.8\%, with an average drop of 51.9\% across all datasets. 

\textbf{Impact of the structural component.}
Similarly, \systR\ (w/o struct.) presents the results of our method without the structural component, i.e., it includes the execution of the factual component only, in one cycle. Despite the fact that running the factual component first yields better results than the reverse order, yet, running the factual component alone, yields low-quality results; usually worse than running the structural component only. This observation does not hold for the \DNine\ and \DTen\ datasets, though, for the reasons described earlier, in Section~\ref{ssec:dataset_analysis}. This means that by running the factual component first, we exploit the two components better, compared to running the structural component first. This observation is also supported by the results presented in Table~\ref{tab:reciprocity_rrea}.

\vspace{-.5cm}
\subsection{Lessons Learned}\label{ssec:lessons}
The highlights of our experiments are summarized below: 
\begin{itemize}
    \item Both structural and factual information play an important role in EA. Our ablation study clearly shows that when we remove the structural component or the factual component, the effectiveness drops significantly. We take advantage of both matching evidences, showing a good generalization among  datasets with diverse structural and semantic heterogeneity.
    \item This important role of both structural and factual information is also supported by the results obtained by the methods that only focus on one side. Structure-based methods perform badly on dataset with high structural heterogeneity while Language Model-based methods perform badly on dataset with high semantic heterogeneity. By considering both cases, \syst\ manages to perform very well on every evaluated dataset.
    \item The reciprocity filter is able to identify matches with high confidence, as also highlighted in~\cite{DBLP:conf/edbt/Efthymiou0SC19,DBLP:conf/cikm/MaoWXWL20}. Table~\ref{tab:reciprocity_rrea} shows that most of the matches returned by \syst \ are due to this filter, with 100\% precision in almost all cases. 
    \item In the whole \syst \ pipeline, it is more beneficial to execute the factual component first (see Table \ref{tab:reciprocity_rrea}). Moreover, using the factual component alone yields bad results, compared to using the structural component alone. The largest improvement in \syst \ effectiveness due to the newly generated matches is observed in the first cycle, while H@1 is not dropping in subsequent cycles (Figure~\ref{fig:perf_cycle}). 
    \item Despite the high number of unnamed nodes along with the high semantic heterogeneity of textual information in multiple languages in multilingual datasets, these datasets may be considered as the easiest datasets, by showing no particularly strong heterogeneities.
    \item Overall, \syst \ is able to adapt to the different degrees of semantic and structural heterogeneity exhibited by real KGs, as seen in Table~\ref{tab:reciprocity_rrea} by the different contribution of the two components across datasets. Intuitively, one component can make up for the lost matches of the other component, leading to high adaptability of our method, enhancing its overall performance.
\end{itemize}
\section{Conclusion}\label{sec:conclusion}

In this paper, we have introduced \syst, an open-source, hybrid EA method that can be build on top of a structural approach. The combination of both factual and structural component allow to cope with datasets of high heterogeneity, with one model capturing structural and one model capturing factual evidence of alignment. We have shown experimentally that \syst\ systematically outperforms state-of-the-art baseline methods, including BERT-INT, Knowformer and RREA, by as far as 29\% Hits@1, obtaining almost perfect scores on every datasets. 

Our choice of two separate models for tackling semantic and structural heterogeneity of KGs can also be considered as an important step towards the need to explain matching decisions~\cite{DBLP:journals/pvldb/TeofiliFKMS22,DBLP:conf/icde/TeofiliFKMMS22,DBLP:conf/sigmod/PradhanLGS22} at different granularity levels to debug, refine, and validate the alignment process: (i) \emph{coarse-grained}: it makes it easy to see the provenance of the returned matches by the structural vs the factual component. (ii) \emph{fine-grained}: the factual model could highlight the most influential attributes in finding a match (e.g., birthDate, foaf:name etc), while the structural model could report the most influential relations that support the matches (e.g., both are connected to similar subgraphs).


Assessing the impact of structural diversity in EA is only a first step in tackling matching bias of KGs. There is strong evidence that entities with rich and high-quality factual information are more likely to be correctly matched, opening a new research work related to group fairness for EA. We plan to extend our graph sampling method \cite{DBLP:conf/esws/FanourakisECKPS23} to also consider the diversity of factual information (i.e., literal values) between two KGs and assess the robustness of EA methods.

\begin{acknowledgements}
This work has received funding from the Hellenic Foundation for Research and Innovation (HFRI) and the General Secretariat for Research and Technology (GSRT), under grant agreement No 969.
\end{acknowledgements}

\bibliographystyle{spmpsci}      
\bibliography{biblio}   

\newpage
\appendix
\section{Appendix A}
\label{appendix}

In this appendix, we provide a robustness analysis of \systR~and \systK, assessing the robustness of our semi-supervised method in newly generated matching pairs. In addition, for completeness purposes, we extend our reciprocity filter analysis of Section~\ref{ssec:results} including results from \systK, with also an efficiency analysis of both \systR~and \systK~and an ablation study showing the impact of each \systK~ component in its effectiveness improvement. Last but not least, we present an ablation study on different initialization techniques of \systR.

\vspace{-0.5cm}
\subsection{Robustness of semi-supervision}
\label{Appendix_robust}
\begin{figure}[ht]
\centering
\begin{subfigure}{\linewidth}
    \centering
    \includegraphics[width=1.0\linewidth]{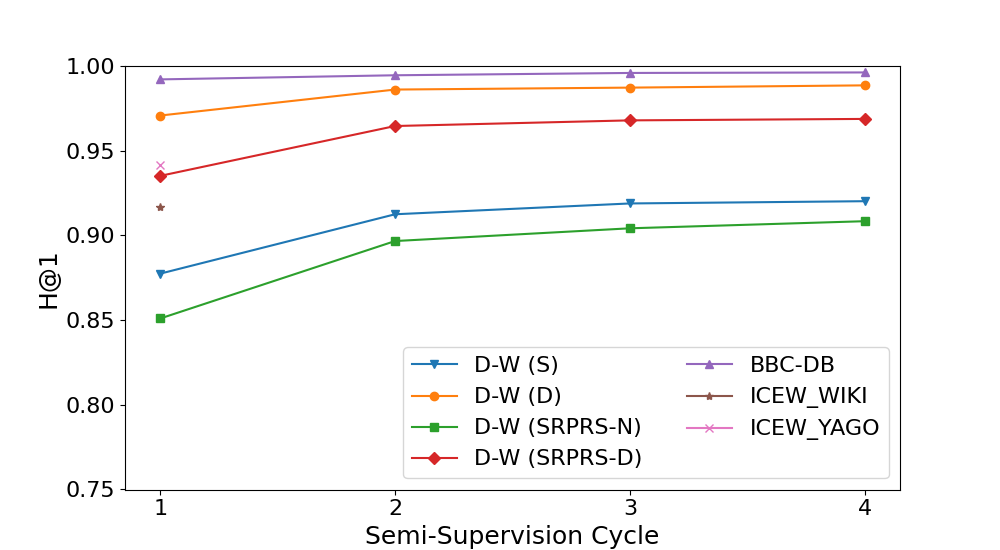}
    \caption{\systK}
    \label{fig:HybEAK_perf_cycle}
\end{subfigure}

\begin{subfigure}{\linewidth}
    \centering
    \includegraphics[width=0.85\linewidth]{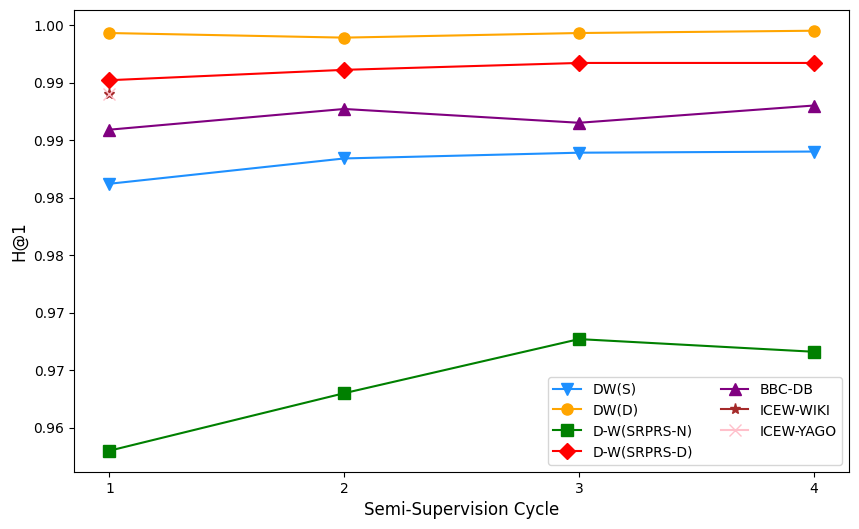}
    \caption{\systR.}
    \label{fig:HybEAR_perf_cycle}
\end{subfigure}
\caption{Comparison of performance per cycle.}
\label{fig:perf_cycle}
\end{figure}

In Figure~\ref{fig:perf_cycle}, we demonstrate H@1 of \systK \ (Figure~\ref{fig:HybEAK_perf_cycle}) and \systR \ (Figure~\ref{fig:HybEAR_perf_cycle}) in each semi-supervision cycle, assessing the robustness of our semi-supervised method to the newly generated matching pairs of each cycle. As we can see from the figure, \systK \ is improved through the cycles in all datasets. The largest improvement is observed from the first to the second cycle, while in the next cycles, the improvement is substantially smaller, due to the rapidly decreasing number of newly generated matching pairs. H@1 is not dropping through cycles, which means that the newly generated matches do not introduce noise. In Figure~\ref{fig:HybEAR_perf_cycle} we demonstrate H@1 of \systR \ , in each semi-supervision cycle, assessing the robustness of our semi-supervised method to the newly generated matching pairs of each cycle. As we can see from the figure, \systR \ is improved through the cycles in mainly more datasets. However, we observe a drop in certain cycles for \systR \ , which could be attributed to the fact that the initial performance is already very high and leave only small space for improvement.  Although certain cycles show slight drops, \systR \ continues to improve from the first to the last cycle, indicating that the method remains effective throughout the process. In both figures, \DNine~and \DTen~are two points, since they only run in one semi-supervision cycle.

\vspace{-0.5cm}
\subsection{Reciprocity filter analysis for \systK}
\label{Appendix_reciprocity}
\begin{table*}[t]
    \centering
    \caption{Cumulative Precision (Pr), Recall (Re), and F1-score (F1) results of the reciprocity filter per dataset and per component, over all iterations, for \systK. Separate evaluation for structural component ran first and for structural component ran first. Maximum recall (max) of the second component is 1 - Re of the first component.}
    \label{tab:reciprocity}
    \begin{tabular}{l|l|rrrr@{\hskip 1cm}|l|rrrr}
         \textbf{Dataset} & \textbf{structural first}  & \textbf{Pr} & \textbf{Re} & (max) & \textbf{F1} & \textbf{factual first} & \textbf{Pr} & \textbf{Re} & (max) & \textbf{F1} \\ \hline
         
         \multirow{3}{*}{\textbf{\DOne}} 
         & structural & 1.0 & 0.75 &        & 0.86 & fact.   & 1.0 & 0.54 &        & 0.70 \\ 
         & factual   & 1.0 & 0.10 & (0.15) & 0.18 & struct. & 1.0 & 0.28 & (0.46) & 0.44 \\
         \cline{2-11}
         & \textbf{cumulative} & 1.0 & 0.85 &  & 0.92 & \textbf{cumulative} & 1.0 & 0.82 &  & 0.90 \\ 
         \hline
         \multirow{3}{*}{\textbf{\DTwo}} 
         & structural & 1.0 & 0.94 &        & 0.97 & fact.   & 1.0 & 0.60 &        & 0.75 \\ 
         & factual   & 1.0 & 0.03 & (0.06) & 0.06 & struct. & 1.0 & 0.35 & (0.40) & 0.51 \\ 
         \cline{2-11}
         & \textbf{cumulative} & 1.0 & 0.97 &  & 0.99 & \textbf{cumulative} & 1.0 & 0.95 &  & 0.98 \\ 
         \hline
         \multirow{3}{*}{\textbf{\DThree}} 
         & structural & 1.0 & 0.60 &        & 0.75 & fact.   & 1.0 & 0.52 &        & 0.68 \\ 
         & factual   & 1.0 & 0.26 & (0.40) & 0.41 & struct. & 1.0 & 0.31 & (0.48) & 0.48 \\
         \cline{2-11}
         & \textbf{cumulative} & 1.0 & 0.86 &  & 0.93  & \textbf{cumulative} & 1.0 & 0.83 &  & 0.91 \\ 
         \hline
         \multirow{3}{*}{\textbf{\DFour}} 
         & structural & 1.0 & 0.88 &        & 0.94 & fact.   & 1.0 & 0.44 &        & 0.61 \\
         & factual   & 1.0 & 0.05 & (0.12) & 0.10 & struct. & 1.0 & 0.48 & (0.56) & 0.65 \\ 
         \cline{2-11}
         & \textbf{cumulative} & 1.0 & 0.93 &  & 0.96 & \textbf{cumulative} & 1.0 & 0.92 &  & 0.96 \\ 
         \hline
         \multirow{3}{*}{\textbf{\DFive}}
         & structural & 1.0 & 0.97 &        & 0.98 & fact.   & 1.0 & 0.73 &        & 0.85 \\
         & factual   & 1.0 & 0.01 & (0.03) & 0.02 & struct. & 1.0 & 0.22 & (0.27) & 0.36 \\
         \cline{2-11}
         & \textbf{cumulative} & 1.0 & 0.98 &  & 0.99  & \textbf{cumulative} & 1.0 & 0.95 &  & 0.97 \\ 
        \hline
        \multirow{3}{*}{\textbf{\DNine}}
         & structural & 1.0 & 0.05 &        & 0.11 & fact.   & 0.99 & 0.91 &  & 0.95 \\
         & factual  & 0.99 & 0.85 & (0.95) & 0.92 & struct. & 1.0 & 0.002 & (0.09) & 0.005 \\
         \cline{2-11}
         & \textbf{cumulative} & 0.99 & 0.91 &  & 0.95 & \textbf{cumulative} & 0.99 & 0.91 & & 0.95 \\ 
         \hline
         \multirow{3}{*}{\textbf{\DTen}}
         & structural & 1.0 & 0.01 & & 0.02 & fact.   & 1.0 & 0.93 &   & 0.96 \\
         & factual   & 1.0 & 0.92 & (0.99) & 0.96 & struct. & 1.0 & 0.002 & (0.07) & 0.002 \\
         \cline{2-11}
         & \textbf{cumulative} & 1.0 & 0.94 &  & 0.96 & \textbf{cumulative} & 1.0 & 0.94 &  & 0.96 \\ 
         \hline
    \end{tabular}
\end{table*}

In Table~\ref{tab:reciprocity}, we extend the reciprocity filter analysis of Section~\ref{ssec:results}, on \systK. More precisely, as described in the corresponding section, it is essential to note that the order of the components is determining the results of the table, since the second component is only called for the pairs that the first component could not match with the reciprocity filter. In other words, the highest achievable recall by the second component is 1 minus the recall of the first component, expressed by the (max) column. 

As in Table~\ref{tab:reciprocity_rrea}, the left half of the table shows the results obtained by running the structural component first and the right half of the table shows the results obtained by running the factual component first (i.e., the default setting). In both cases, we also report the maximum achievable recall (max) by the second component. By comparing the cumulative F1 scores for the same dataset in the two variations, we notice that by executing the structural component first, we achieve better results than executing the factual component first, in all datasets. We also observe that the contribution of the factual component, when executed second, is very limited, since the effectiveness of the structural component does not allow much room for improvement. On the other hand, when executing the factual component first, the contribution of the two components is more balanced. Note that those results only reflect the effectiveness of the reciprocity filter and they do not reflect the overall evaluation of \systK, although the reciprocity filter largely impacts the overall effectiveness of \systK.

Another very important observation comes up from the comparison of the H@1 scores of \systK\ and \systK\ (struct. first) in Table~\ref{tab:ablation_knowformer} and the Recall results (since Precision is almost always 1.0) shown in the factual first and structural first parts of Table~\ref{tab:reciprocity}, respectively. We notice that the vast majority of matches identified by \systK\ come from the reciprocity filter and only a very small number of matches are left to be decided by the ranking of the most likely matches returned by the last component. 

In more detail, we observe the following: 
\begin{itemize}
    \item \DOne: When executing the structural component first, the cumulative recall from the reciprocity filter is 0.85 and H@1 of \systK\ (struct. first) is 0.86, i.e., a 1\% total difference. 
    When executing the factual component first, the cumulative recall from the reciprocity filter is 0.82, but H@1 of \systK\ is 0.92, i.e., a 10\% total difference.

    \item \DTwo: When executing the structural component first, the cumulative recall from the reciprocity filter is 0.97 and H@1 of \systK\ (struct. first) is also 0.97, i.e., the number of correct matches identified outside the reciprocity filter is negligible. 
    When executing the factual component first, the cumulative recall from the reciprocity filter is 0.95, but H@1 of \systK\ is 0.98, i.e., a 2\% total increase.

    \item \DThree: When executing the structural component first, the cumulative recall from the reciprocity filter is 0.86 and H@1 of \systK\ (struct. first) is also 0.86, again with negligible contribution of non-reciprocity-filter suggested matches. 
    When executing the factual component first, the cumulative recall from the reciprocity filter is 0.83, but H@1 of \systK\ is 0.90, i.e., a 7\% increase.

    \item \DFour: With the structural component first, the cumulative recall from the reciprocity filter is 0.93 and H@1 of \systK\ (struct. first) is 0.93, once again a negigible contribution of the matches not provided by the reciprocity filter. 
    When executing the factual component first, the cumulative recall from the reciprocity filter is 0.92, but H@1 of \systK\ is 0.96, i.e., a 4\% difference.

    \item \DFive: The cumulative recall from the reciprocity filter when executing the structural component first is 0.98, while H@1 of \systK\ (struct. first) is 0.98, once again showing no difference.
    When executing the factual component first, the cumulative recall from the reciprocity filter is 0.95, but H@1 of \systK\ is 0.99, i.e., an almost 4\% total difference.
    
    \item \DNine: The cumulative recall obtained by the reciprocity filter when using the structural component first is 0.91, and H@1 of \systK\ (struct. first) is 0.98. This is the first time we observe a significant difference between the recall of the reciprocity filter and the performance of the model when using the structural component first.
    When executing the factual component first, the cumulative recall from the reciprocity filter is 0.91, while H@1 of \systK\ is also 0.91. This is the first time we observe a negligible contribution between the recall of the reciprocity filter and the performance of the model when using the factual component first. It is also the first that we see the precision not be 1.0, but 0.99, proposing 1\% wrong matching pairs.
    
    \item \DTen: The cumulative recall obtained by the reciprocity filter when using the structural component first is 0.94, and H@1 of \systK\ (struct. first) is 0.95, i.e. a 1\% difference.
    When executing the factual component first, the cumulative recall from the reciprocity filter is 0.94, while H@1 of \systK\ is 0.94, i.e., once again a negigible contribution when using factual first. In addition, this dataset is another case where the precision is 0.99 istead of 1.0.
\end{itemize}

From the above findings, we can observe that the effectiveness of the reciprocity filter is on par with the results exhibited by \systK: both are higher when using the factual component first, except from the datasets \DNine\ and \DTen. It is also interesting to note that the highest differences observed between the cumulative recall of the reciprocity filter and H@1 are on \DOne\, \DThree\, \DFour\ which are the datasets that showed the most important semantic heterogeneity in the dataset analysis section. It can thus be explained by the fact that one of the component has more difficulties to correctly align, leading to a significant difference between the reciprocity filter and the performance.

\begin{table*}[t]
\centering
\caption{Efficiency results of \systK.}
\label{tab:efficiency_HybEAK}
\begin{tabular}{|l|ll|ll|l|l|l|l|}
\hline
\multirow{2}{*}{} & \multicolumn{2}{l|}{Mean time per epoch (s)} & \multicolumn{2}{l|}{Total number of epochs} &\multirow{2}{*}{\shortstack{Number of semi- \\  supervision cycles}} & \multirow{2}{*}{\shortstack{Total time \\factual (s)}} & \multirow{2}{*}{\shortstack{Total time \\ structural (s)}} & \multirow{2}{*}{Total time (s)} \\ \cline{2-5}
                  & Factual & Structural& Factual & Structural&& &&\\ \hline

\DOne & 65.77& 32.81 & 230 & 325 & 4 & 15,127 & 10,663 & 25,790\\ \hline
\DTwo & 59.74 & 62.70 & 220 & 205 & 4 & 13,142 & 12,853 & 25,996\\ \hline
\DThree & 57.54 & 31.88 & 410 & 260 & 4 & 23,591 & 8,288 & 31,880\\ \hline
\DFour  & 53.55 & 57.77 & 300 & 235 & 4 &16,065  & 13,575 & 29,640\\ \hline
\DFive  & 57.28  & 23.17 & 140 & 125 & 4 & 8,019 & 2,896 & 10,915\\ \hline
ICEWS-WIKI  & 4.05 & 1,455 & 30 & 15 & 1 & 122 & 21,836 & 21,958 \\ \hline
ICEWS-YAGO  & 19.12  & 1,873 & 30 & 20 & 1 & 574 & 37,475 & 38,049 \\ \hline

\end{tabular}
\end{table*}
\begin{table*}[t]
\centering
\caption{Efficiency results of \systR.}
\label{tab:efficiency_HybEAR}


\begin{tabular}{|l|ll|ll|l|l|l|l|}
\hline
\multirow{2}{*}{} & \multicolumn{2}{l|}{Mean time per epoch (s)} & \multicolumn{2}{l|}{Total number of epochs} &\multirow{2}{*}{\shortstack{Number of semi- \\  supervision cycles}} & \multirow{2}{*}{\shortstack{Total time \\factual (s)}} & \multirow{2}{*}{\shortstack{Total time \\ structural (s)}} & \multirow{2}{*}{Total time (s)} \\ \cline{2-5}
                  & Factual & Structural& Factual & Structural&& &&\\ \hline

\DOne & 65.77& 0.22 & 240 & 4,800 & 4 & 15,785 & 1,056 & 16,841 \\ \hline
\DTwo & 59.74 & 0.30 & 200 & 4,800 & 4 & 11,948 & 1,440 & 13,388 \\ \hline
\DThree & 57.54 & 0.21 & 435 & 4,800 & 4 & 25,030& 1,008& 26,038 \\ \hline
\DFour  & 53.55 & 0.28 & 250 & 4,800 & 4 &13,388 & 1,344  & 14,732  \\ \hline
\DFive  & 57.28  & 0.13& 110 & 4,800 &4 & 6,360 & 624 & 6,924 \\ \hline
ICEWS-WIKI   & 4.05 & 1.12 & 30 & 1,200 & 1 & 122 & 1,344 & 1,466 \\ \hline
ICEWS-YAGO & 19.12 & 1.46 & 30 & 1,200 &1 &574&1,752& 2,326 \\ \hline
\end{tabular}
\end{table*}

\subsection{Efficiency analysis}
In Tables~\ref{tab:efficiency_HybEAK} and~\ref{tab:efficiency_HybEAR}, we present an efficiency analysis of both \systK\ and \systR\ building blocks respectively, providing the mean time per epoch for both factual and structural components, while also the total number of epochs needed by each component per dataset and the number of the semi-supervised cycles. In addition, appart from the total time that each component requires for training, we also provide the total training time that \systR\ and \systK\ needed for each dataset.

\textbf{Efficiency of building blocks} We observe that despite the fact that the factual component brings huge improvement in our model (as we analyze later in Appendix~\ref{sec:ablation_knowformer} and previously in Section~\ref{sec: ablation_study}), it is less efficient that the structural component in the vast majority of the datasets in both \systR\ and \systK (e.g., the factual component of \systK\ in \DThree\ needs 23,591s, in contrast to the structural one which needs 8,288). This is not the case for the highly structural heterogeneous \DNine\ and \DTen\, burdening the structural component in terms of the mean time required per epoch, leading to higher total training time (e.g., 1,455s for each epoch and 21,836s total training time for \DNine\ and \systK). Also, it is worth mentioning that the reciprocity filter (another building block of HybEA), despite the huge improvenent in terms of effectiveness that brings, it requires negligible time, that is near to zero.

Despite the large number of epochs that the structural component of \systR\ (i.e., RREA(basic)) requires (1,200 per semi-supervised cycle), the total time it needs is less than the structural component of \systK\ (i.e., Knowformer) with two order of magnitude less mean time per epoch (e.g., for \DOne\ the structural component of \systR\ needs 0.22s, while the one of \systK\ needs 32.81s). On the other hand, the higher effectiveness of RREA(basic) compared to Knowformer (as shown in Table~\ref{tab:effectiveness}), leads to less overhead in terms of time and number of epochs of \systR\ factual model, compared to the one of \systK\ in most datasets (e.g., the factual component of \systR\ for \DTwo\ needs 200 epochs and 11,948s, while the one of \systK\ needs 220 epochs and 13,142s). \DOne\ and \DThree\ are the exception datasets, since they exhibit higher semantic heterogeneity, requiring bigger effort by the factual component. For \DNine\ and \DTen\ the training time that was required by the factual model remains the same for both \systR\ and \syst, since it only ran in one semi-supervised cycle with negligible number of newly generated matching pars from the reciprocity filter.

\textbf{Efficiency per dataset} In both tables, we observe that while most of the datasets need four semi-supervision cycles to train, \DNine\ and \DTen\ need only one, since the newly generated matching pairs are limited, consequently the low number of total epochs in both \systR\ and \systK and training time needed by each component individually and in total in \systR\. In addition, since \DFive\ is a dataset with both low factual and structural heterogeneity compared to the other datasets, it needs lower time in total, but also for each component individually, for both \systR\ and \systK. Last but not least, we observe that the dense versions \DTwo\ and \DFour\ of \DOne\ and \DThree, require more effort by the structural component due to the more structural information, leading to higher individual and total training time in both \systR\ and \systK, compared to the sparse versions.

\textbf{Efficiency per method}
\systR\ is more efficient than \systK\ in terms of the total time required for training, while also it outperforms it in almost all the datasets in terms of effectiveness (see Table~\ref{tab:effectiveness}). This is due to the high efficient and performant structural component of \systR(RREA), that outperforms Knowformer of \systK\ in both efficiency and effectiveness.

\begin{table*}[t]
    \centering
    \caption{Ablation study for \systK.}
    \label{tab:ablation_knowformer}
    \begin{tabular}{ll|rrrrrrr}
        \textbf{Method} & \textbf{Metric} & \textbf{\DOne} & \textbf{\DTwo} & \textbf{\DThree} & \textbf{\DFour} & \textbf{\DFive} & \textbf{ICEWS-WIKI} & \textbf{ICEWS-YAGO} \\ \hline
        
        \multirow{3}{*}{\textbf{\systK}} 
         & H@1 & \textbf{0.920} & \textbf{0.988}  & \textbf{0.908}  & \textbf{0.968}  & \textbf{0.996} & 0.916 & 0.941 \\
         & H@10  & \textbf{0.969} & \textbf{0.997} & \textbf{0.954} & \textbf{0.987}  & \textbf{0.997} & 0.938 & 0.944\\
         & MRR & \textbf{0.93} & \textbf{0.99} & \textbf{0.92} & \textbf{0.97} & \textbf{0.99} & 0.92 & 0.94\\ \hline 
         
          \multirow{3}{*}{\textbf{\shortstack[l]{\systK \\ (basic)}}}   
         & H@1  & 0.877 & 0.970 & 0.850 & 0.935 &  0.992 & 0.916 & 0.941\\
         & H@10 & 0.962 & 0.995 & 0.936 & 0.981 & 0.996 &  0.938 & 0.944\\
         & MRR  & 0.90 & 0.97 & 0.88 & 0.95 & 0.99 & 0.92 & 0.94\\ \hline
         
         \multirow{3}{*}{\textbf{\shortstack[l]{\systK \\ (basic; str. first)}}}   
         & H@1 & 0.820 & 0.947  & 0.827  & 0.913 & 0.977 & \textbf{0.985} & \textbf{0.952}\\
         & H@10  & 0.837  & 0.955 & 0.855 & 0.919 & 0.978 & \textbf{0.995} & \textbf{0.976}\\
         & MRR & 0.83 & 0.95 & 0.84 & 0.91 & \underline{0.98} & \textbf{0.98} & \textbf{0.96}\\ \hline

                \multirow{3}{*}{\textbf{\shortstack[l]{\systK \\ (struct. first)}}}
        
         & H@1 & 0.861 & 0.971  & 0.863 & 0.938 & 0.984 & \textbf{0.985} & \textbf{0.952}\\
         & H@10  & 0.872 & 0.974 & 0.881 & 0.942 & 0.985 & \textbf{0.995} & \textbf{0.976}\\
         & MRR & 0.86 & 0.97 & 0.87 & 0.94 & 0.99 & \textbf{0.98} & \textbf{0.96}\\ \hline

         \multirow{3}{*}{\textbf{\shortstack[l]{\systK \\ (w/o fact.)}}} 
         & H@1 & 0.753 & 0.917  & 0.603  & 0.873 & 0.974 & 0.073 & 0.019\\
         & H@10  & 0.936 & 0.987 & 0.815 & 0.970 & 0.995 & 0.275 & 0.095\\
         & MRR & 0.82 & 0.94 & 0.67 & 0.90 & \underline{0.98} & 0.13 & 0.04\\ \hline

         \multirow{3}{*}{\textbf{\shortstack[l]{\systK \\ (w/o struct.)}}} 
         & H@1 & 0.558 & 0.621  & 0.513  & 0.444 & 0.748 & 0.926 & 0.948\\
         & H@10  & 0.612 & 0.672 & 0.592 & 0.488 & 0.769 & 0.974 & 0.975\\
         & MRR & 0.79 & 0.82 & 0.70 & 0.70 & 0.91 & 0.94 & 0.96\\ \hline

    \end{tabular}
\end{table*}

\begin{table*}[h]
    \centering
    \caption{Ablation study on \systR\ initialization.}
    \label{tab:performance_hybEAR}
    \begin{tabular}{ll|rrrrrrr}
        \textbf{Method} & \textbf{Metric} & \textbf{\DOne} & \textbf{\DTwo} & \textbf{\DThree} & \textbf{\DFour} & \textbf{\DFive} & \textbf{ICEWS-WIKI} & \textbf{ICEWS-YAGO} \\ \hline
         
        \multirow{3}{*}{\textbf{\shortstack[l]{\systR \\ with random \\ initialization}}}
     & H@1 & 0.989 & \textbf{1.000}  & 0.972  & \textbf{0.997}  & 0.993 & \textbf{0.994} & \textbf{0.994} \\
     & H@10  & 0.997 & \textbf{1.000} & 0.992 & \textbf{1.000}  & \textbf{1.000} & \textbf{0.997} & 0.997 \\
     & MRR & \textbf{0.99} & \textbf{1.00} & \textbf{0.98} & \textbf{1.00} & \textbf{1.00} & \textbf{1.00} & \textbf{1.00} \\ \hline
         
        \multirow{3}{*}{\textbf{\shortstack[l]{\systR \\ with SentenceBert \\ initialization}}}
         & H@1 & \textbf{0.990} & 0.999 & \textbf{0.974} & 0.996 & \textbf{0.994} & \textbf{0.994} & \textbf{0.994}\\
         & H@10  & \textbf{0.998} & \textbf{1.000} & \textbf{0.993} & \textbf{1.000} & \textbf{1.000} & \textbf{0.997} & \textbf{0.998}\\
         & MRR & \textbf{0.99} & \textbf{1.00} & \textbf{0.98} & \textbf{1.00} & \textbf{1.00} & 0.99 & \textbf{1.00}\\ \hline

    \end{tabular}
\end{table*}

\vspace{-0.5cm}
\subsection{Ablation study of \systK~components}
\label{sec:ablation_knowformer}

We conduct an ablation study, showing that by changing any single component of \systK, the entire performance drops (Table~\ref{tab:ablation_knowformer}). This demonstrates that all of our architectural choices have a positive impact on the effectiveness of \systK.

\textbf{Impact of component ordering.} We experiment with different orders of execution for the structural and factual components, as presented in row \systK\ (struct. first) of Table~\ref{tab:ablation_knowformer}. We observe that we obtain different results, with winner in the vast majority of the datasets, being running the factual component first (\systK), that was finally chosen as our default configuration.

\textbf{Impact of the semi-supervision step.} Table~\ref{tab:ablation_knowformer} reports the results of our method at the end of the first cycle, when running the structural component first, denoted as \systK\ (basic; str.first), as well as when running the factual component first, denoted as \systK\ (basic). Compared to the results of the full pipeline, with the corresponding order of components, the basic version always yields lower or equivalent results. The difference between the basic version of \systK\ and the full version of \systK\ is bigger than the corresponding of Table~\ref{tab:ablation_RREA}. The difference between the basic and the full version is similar when comparing the results of \systK\ (basic; str. first) with \systK\ (struct. first). It is worth noticing that the basic version of \systK\ already outperforms all the state-of-the-art baseline methods. 

\textbf{Impact of the factual component.} 
The row \systK\ (w/o fact.) skips the factual component and essentially runs the structural component of \systK\ only (i.e., Knowformer) using SentenceBERT initialization and enriched with the reciprocity filter. This experiment involves only one cycle, since re-training the same model without new pairs suggested from the other model, yields the same results in subsequent cycles. Although the reciprocity filter results show perfect precision in almost all cases (in Table~\ref{tab:reciprocity}), the results of \systK\ (w/o fact.) in Table~\ref{tab:ablation_knowformer} are not identical with the results of Knowformer in Table~\ref{tab:effectiveness}, since Knowformer does not use SentenceBERT initialization. Instead, \systK\ (w/o fact.) outperforms Knowformer, showing that the SentenceBERT initialization of the entity embeddings brings big improvement. We observe that \emph{the factual component has a huge impact} on the results, since, when we remove it, the H@1 scores drop by 2.2\% to even 84.6\%.

\textbf{Impact of the structural component.}
Similarly, \systK\ (w/o struct.) presents the results of our method without the structural component, i.e., it includes the execution of the factual component only, in one cycle. Despite the fact that running the factual component first yields better results than the reverse order, yet, running the factual component alone, yields low-quality results; usually worse than running the structural component only. This observation does not hold for the \DNine\ and \DTen\ datasets, though, for the reasons described earlier, in Section~\ref{ssec:dataset_analysis}. This means that by running the factual component first, we exploit the two components better, compared to running the structural component first.

\vspace{-0.5cm}
\subsection{Ablation study on \systR~initialization}
\label{Appendix_initialization}

We analyze the influence of different types of entity initialization strategies on \systR \ performance. Specifically, we compare random initialization with Sentence-BERT initialization. As shown in Table~\ref{tab:performance_hybEAR}, the results reveal that the choice of initialization has little to no impact on the performance of \systR. The H@1, H@10, and MRR scores remain almost identical across all datasets, with only negligible variations. The largest difference is observed on the \DThree dataset, where there is a 0.2\% difference in H@1. This proves that \systR \ efficiently learns robust entity embeddings, making it less dependent on pre-trained embeddings. Given that \systR \ already captures most alignment cues effectively, the additional information from Sentence-BERT initialization has only small impact on performance. Based on these results, we opted to keep random initialization for \systR, as it gives the same high performance while avoiding the extra computational cost of Sentence-BERT embeddings.


\end{document}